\documentclass[%
 reprint,
 amsmath,amssymb,
 aps,
]{revtex4-2}

\usepackage{graphicx}
\usepackage{dcolumn}
\usepackage{bm}
\usepackage{subfigure}
\usepackage{braket}
\usepackage{xcolor}
\usepackage{tabularx}

\newcommand{\comment}[1]{{}}

\begin{document}

\preprint{APS/123-QED}

\title{Quantifying Unknown Entanglement by Neural Networks}

 \author{Xiaodie Lin}
 \author{Zhenyu Chen}
 \author{Zhaohui Wei}\email{Email: weizhaohui@gmail.com}
 \affiliation{
  Center for Quantum Information, Institute for Interdisciplinary Information Sciences, Tsinghua University, Beijing 100084, China
 }

\begin{abstract}
 Quantum entanglement plays a crucial role in quantum information processing tasks and quantum mechanics, hence quantifying unknown entanglement is a fundamental task. However, this is also challenging, as entanglement cannot be measured by any observables directly. In this paper, we train neural networks to quantify unknown entanglement, where the input features of neural networks are the {outcome} statistics data produced by locally measuring target quantum states, and the training labels are well-chosen quantities. For bipartite quantum states, this quantity is coherent information, which is a lower bound for the entanglement of formation and the entanglement of distillation. For multipartite quantum states, we choose this quantity as the geometric measure of entanglement. It turns out that the neural networks we train have very good performance in quantifying unknown quantum states, and can beat previous approaches like semi-device-independent protocols for this problem easily in both precision and application range. We also observe a surprising phenomenon that on quantum states with stronger quantum nonlocality, the neural networks tend to have better performance, though we do not provide them any knowledge on quantum nonlocality.
\end{abstract}

\maketitle

\section{Introduction}

It is well-known that quantum entanglement plays a central role in designing quantum information processing tasks and understanding quantum mechanics. As a result, characterizing unknown quantum entanglement is a fundamental problem. However, as entanglement cannot be measured directly using any physical observables, this is a very challenging task that has attracted a lot of attentions.

Two most popular approaches adopted by quantum experimentalists for this problem are quantum tomography and entanglement witness~\cite{chuang1997prescription,poyatos1997complete,guhne2009entanglement,rosset2012imperfect}, though both of them suffer from obvious drawbacks. Indeed, quantum tomography extracts the full information of target quantum systems, which allows us to look into the underlying entanglement. However, to achieve this exponential amount of resources is needed, thus tomography is not a realistic option even in the realm of noisy, intermediate scale quantum (NISQ) era. On the other hand, though the idea of entanglement witness is much more efficient, it needs much nontrivial prior knowledge on target systems, and usually can only provide an unreliable yes or no answer on whether entanglement exists~\cite{rosset2012imperfect}. In order to overcome the difficulties of these two approaches, {device-independent protocols have been proposed to certify the existence of entanglement~\cite{collins2002nonsep,bancal2011device,pal2011multisetting,murta2016quantum,baccari2017efficient,tavakoli2018semi,zwerger2019device}}, where all conclusions are drawn based only on observed quantum nonlocality, and thus are reliable.

Meanwhile, as quantum computing becomes more and more important from the viewpoint of engineering, how to \emph{quantify} unknown entanglement, an even harder task than only certifying the existence, has also been an important and urgent issue. However, the {quantification} of entanglement is a very complicated problem when it comes to mixed quantum states and multipartite quantum states~\cite{horodecki2009quantum}, where quite a lot of reasonable measures have been proposed, and most of them are very hard to calculate even if the full information of target quantum states is given, not to mention experimentally {quantifying unknown multipartite entanglement}.

Among these measures of entanglement, the quantum R\'{e}nyi entropies and logarithmic negativity are relatively easier to handle~\cite{horodecki2009quantum,zyczkowski1998volume,eisert1999comparison}, thus have been extensively studied in recent years. Indeed, some interesting protocols have been proposed to characterize these two measures by analyzing the data produced by well-designed measurements~\cite{daley2012measuring,abanin2012measuring,brydges2019probing,zhou2020single}. Particularly, machine learning tools like neural networks have been introduced to quantity logarithmic negativity, which provides very nice performance~\cite{gray2018machine} (see also {Ref.}\cite{berkovits2018extracting}). However, it should be pointed out that, in addition to the quantum R\'{e}nyi entropies and logarithmic negativity, we have many more important measures of entanglement, say the entanglement of formation and the entanglement of distillation~\cite{horodecki2009quantum}. Undoubtedly, proposing protocols that are able to experimentally quantify these widely-used measures of entanglement is a very important task.

In fact, device-independent protocols have also been utilized to quantify unknown quantum states~\cite{shahandeh2017measurement,rosset2018practical,guo2020measurement,moroder2013device,shahandeh2017measurement,bardyn2009device,kaniewski2016analytic}. Particularly, by defining a concept called nondegenrate Bell inequalities, a semi-device-independent scheme that can lower bound the coherent information for any unknown bipartite quantum states was found~\cite{wei2021analytic}. Note that coherent information is a lower bound for the entanglement of formation and the entanglement of distillation~\cite{cornelio2011entanglement}, implying that this actually provides an approach partially quantifying these two important measures of entanglement. Later, this approach has been generalized to quantify multipartite entanglement by the geometric measure and the relative {entropy} of entanglement~\cite{lin2020quantifying}.

However, the results provided by device-independent protocols like those in Refs.\cite{wei2021analytic,lin2020quantifying} are very conservative and weak. In fact, usually these protocols can provide informative results only when the observed nonlocality is close to maximal {violation}, which seriously restricts their applications. As a result, new schemes that can experimentally quantify these entanglement measures with better performance and wider application ranges have to be explored, which is also the main motivation of the current paper.

In this paper, we apply neural networks to quantify unknown quantum entanglement {by several popular entanglement measures}. Specifically, to train the neural networks, we choose the outcome statistics data produced by locally measuring sample quantum states as features, and choose coherent information for bipartite quantum states, or the geometric measure of entanglement for multipartite quantum states, as sample labels or prediction targets. Note that for bipartite quantum states, coherent information is a lower bound for the entanglement of formation and the entanglement of {distillation,} thus our results can be regarded as partial quantifications of these two important entanglement measures. By running the trained neural networks on unknown test states, we show that they have very good performance in quantifying the involved entanglement. For example, the prediction results beat those of known device-independent protocols for this problem easily in both precision and application range, implying the high potential value of neural {networks} in quantifying unknown quantum entanglement. Furthermore, we observe interesting evidences showing that neural networks have better performance on quantum states with stronger nonlocality, though our training data does not contain any direct information on quantum nonlocality.

\section{Settings and Entanglement Measures}

As mentioned above, the data {features} that neural networks work with {are} the measurement {outcome} statistics produced by local measurements on target quantum states. Suppose we have a physical system shared by $n$ space-separated parties, and the state is $\rho$. Each party $i$, where $i\in\{1,..,n\}$, has a set of measurement devices labelled by $X_i$ and the possible measurement outcomes are labelled by $A_i$. Assume every party randomly chooses a measurement $x_i\in X_i$ to measure their subsystems and record the corresponding outcomes $a_i\in A_i$. After repeating the procedure for many times, they have the joint probability distribution $p(a_1a_2...a_n|x_1x_2...x_n)$, which indicates the probability of obtaining outcomes $a_i\in A_i$ when measurements $x_i\in X_i$ are chosen. Suppose $M_{x_i}^{a_i}$ is the measurement operator with outcome $a_i$ for the measurement device $x_i$ performed by the party $i$, then it holds that
\begin{eqnarray}\label{Correlation}
p(a_1a_2...a_n|x_1x_2...x_n)={\rm Tr}\left(\left(\bigotimes_{i=1}^nM_{x_i}^{a_i}\right)\rho\right).
\end{eqnarray}

Note that {although} our setting is quite similar with Bell experiments, we will not talk about nonlocality and Bell inequalities at most time, though like in Bell experiments we hope that neural networks will reveal us nontrivial information on the amount of entanglement based only on the data $p(a_1a_2...a_n|x_1x_2...x_n)$.

For this, we have to choose proper entanglement measures for our tasks. Before doing that, let us first recall briefly how neural networks, a machine learning model we will utilize in this paper, work. First, we set up a proper mathematical model for the target problem, which usually contains one input layer, zero or more hidden layers, and one output layer, and {every two adjacent layers involve} a lot of parameters. Then we gather many representative sample data with correct labels, called training data, and input them into the model. We hope the outputs of the model are as consistent as possible with the correct labels, for which we constantly adjust the parameters by proper optimization methods, and this process is called training. If we train the model correctly, eventually it will capture crucial properties of the training dataset, and can even predict the labels very precisely on an unknown test dataset.

Accordingly, in our task the outputs of neural networks, or the labels of training dataset, are supposed to be target entanglement measures we are interested in. As a consequence, in order to prepare sufficient representative training data with correct labels, we have to make sure that the corresponding entanglement measures can be computed accurately on all training states. As is well-known, this is a very challenging task. For example, if we choose the entanglement of formation and the entanglement of distillation as our target entanglement measures, this requirement is very hard to satisfy. Therefore, if we insist on our target entanglement measures, we have to make proper compromises.

For bipartite quantum states, a possible way to compromise is to quantify coherent information instead. Coherent information is a fundamental quantity responsible for the capability of transition of quantum information \cite{schumacher1996quantum,lloyd1997capacity}. For a quantum state $\rho$ acting on a composite Hilbert space $\mathcal{H}_A\otimes\mathcal{H}_B$, its coherent information is defined as
\begin{eqnarray}\label{CI}
I_C(\rho)=S(\rho_A)-S(\rho),
\end{eqnarray}
where $\rho_A={\rm Tr}_B(\rho)$ is the subsystem in $\mathcal{H}_A$ and $S(\rho)$ is the Von Neumann entropy of $\rho$. It can be seen that the coherent information is relatively easy to calculate, and most importantly, it turns out that \cite{cornelio2011entanglement}
\begin{eqnarray}\label{CI_lower_bound}
E_f(\rho)\ge E_D(\rho)\ge I_C(\rho),
\end{eqnarray}
where $E_f(\rho)$ is the entanglement of formation of $\rho$ and $E_D(\rho)$ is the distillable entanglement of $\rho$. Therefore, if we choose the prediction target of neural networks as coherent information, the training data should be easy to gather, and if the model has good performance, we can eventually obtain nontrivial information {on} $E_f(\rho)$ and $E_D(\rho)$ for target quantum states.

We now turn to the case of multipartite quantum states. Unfortunately, in this case very few well-known entanglement measures or quantities related to entanglement measures can be calculated accurately, which makes it very hard to prepare representative training data for our neural networks. In this paper, we choose the geometric measure of entanglement (GME) as our target entanglement measure. The reason is that by numerical methods, we can barely manage to calculate the entanglement measure of some pure multipartite quantum states, which serve as the training data.

Recall that the GME is firstly introduced in the setting of bipartite pure states \cite{shimony1995degree}, and then generalized to the multipartite setting \cite{barnum2001monotones}. For a $n$-partite pure state $|\psi\rangle$, the GME is defined as
\begin{eqnarray}\label{GME_Pure}
E_G(|\psi\rangle)=1-G(|\psi\rangle)^2,
\end{eqnarray}
where
\begin{eqnarray}\label{G}
G(|\psi\rangle)=\sup_{|\phi\rangle\in {\rm sep}_n}|\langle\psi|\phi\rangle|,
\end{eqnarray}
and ${\rm sep}_n$ is the set of $n$-partite product pure states. For a mixed state $\rho$, the GME is defined by convex roof construction, which is
\begin{eqnarray}\label{GME_Mixed}
E_G(\rho)=\min_{\rho=\sum_ip_i|\psi_i\rangle\langle\psi_i|}\sum_iE_G(|\psi_i\rangle).
\end{eqnarray}

\section{Quantifying Bipartite Entanglement}

\subsection{Qutrit-qutrit case}

As a warm-up, we first consider quantifying the coherent information of a special class of qutrit-qutrit quantum states, which have the form
\begin{eqnarray}\label{CGLMP_Class}
\rho_{3_{\epsilon}}^{mv}=(1-\epsilon)I_9/9+\epsilon|\psi_3^{mv}\rangle\langle\psi_3^{mv}|,
\end{eqnarray}
where $\epsilon\in[0,1]$ and $|\psi_3^{mv}\rangle=\gamma(|00\rangle+|22\rangle)+\sqrt{1-2\gamma^2}|11\rangle$ with $\gamma\approx0.617$. Note that $|\psi_3^{mv}\rangle$ is the quantum state that violates the Collins-Gisin-Linden-Massar-Popescu (CGLMP) inequality maximally~\cite{collins2002bell,acin2002quantum,chen2006violating,zohren2008maximal}. The reason why we choose this class of states as our first target is that their coherent information can also been quantified using the semi-device-independent approach proposed in Ref.\cite{wei2021analytic}, which provides us a nice chance to compare the performance of these two different approaches.

As mentioned above, the sample features for the neural networks we choose {are} the {outcome} statistics data produced by locally measuring the involved quantum states. Here the local measurements we choose are the ones that maximize the violation of the CGLMP inequality, which can be described by observables $A_a$ and $B_b$ with the eigenvectors
\begin{eqnarray}\label{CGLMP_Measurement1}
|i\rangle_{A,a}=\frac{1}{\sqrt{d}}\sum_{k=0}^{d-1}{\rm exp}\left({\mathbf i}\frac{2\pi}{d}k(i+\alpha_a)\right)|k\rangle_A
\end{eqnarray}
and
\begin{eqnarray}\label{CGLMP_Measurement2}
|j\rangle_{B,b}=\frac{1}{\sqrt{d}}\sum_{l=0}^{d-1}{\rm exp}\left({\mathbf i}\frac{2\pi}{d}l(-j+\beta_b)\right)|l\rangle_B
\end{eqnarray}
respectively~\cite{collins2002bell,barrett2006maximally}, where $a,b\in\{1,2\}$, $i,j\in\{0,1,...,d-1\}$, ${\mathbf i}=\sqrt{-1}$ is the imaginary number, and the phases read $\alpha_1=0,\alpha_2=1/2,\beta_1=1/4$ and $\beta_2=-1/4$. In this subsection, we let $d=3$.

To generate the training dataset, a number of bipartite quantum states of form $(1-\epsilon)I_9/9+\epsilon\rho_0$ are sampled. Here we select $\epsilon$ from 0.2 to 1 at intervals 0.025 and $\rho_0$ as arbitrary quantum state in $\mathcal{H}^3\otimes\mathcal{H}^3$. For each $\epsilon$ not larger than 0.75, 1,000 states are randomly sampled, while for the other values of $\epsilon$, we increase the number of samples to 2,000. Then, the corresponding outcome statistics data are recorded when measuring $A_a$ and $B_b$ locally on the sampled states. Together with the corresponding coherent information as the {labels}, 43,000 instances as training dataset are fed to a 4-hidden-layer {fully connected} neural network with 400, 200, 100, 50 neurons in each layer, respectively.
\begin{figure}[!ht]
	\centering
	\includegraphics[width=0.45\textwidth]{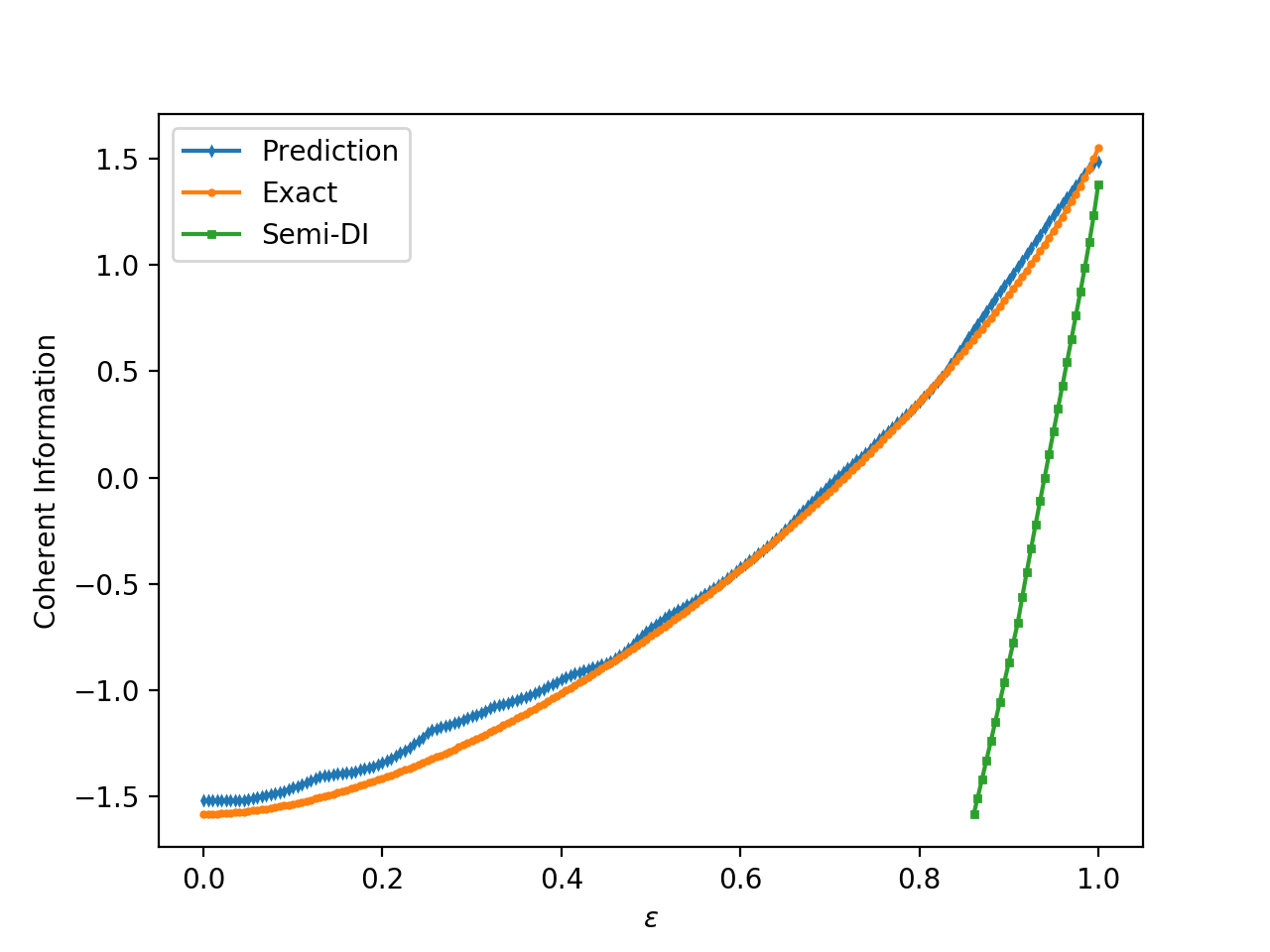}
	\caption{The neural network predictions for the coherent information of $\rho_{3_{\epsilon}}^{mv}$. The blue line and the {orange line} represent the prediction values and the exact values, respectively, where the MSE of these prediction values is 0.0040. In addition, the green line shows the lower bound for the coherent information provided by the semi-device-independent (semi-DI) approach in Ref.\cite{wei2021analytic}, which has a much narrower application range.}
	\label{fig:result_CGLMP}
\end{figure}

After training, we apply our model to predict the coherent information of $\rho_{3_{\epsilon}}^{mv}$ and the test {dataset} is generated by selecting $\epsilon$ from 0 to 1 at intervals 0.005. We evaluate the performance of the model with mean squared error (MSE), which is {defined} as
\begin{eqnarray}\label{MSE}
{\rm MSE}=\frac{1}{N}\sum_{i=1}^{N}(Y_i-\hat{Y}_i)^2,
\end{eqnarray}
where $Y_i$ is the exact value of coherent information, $\hat{Y}_i$ is the predicted value, and $N$ is the {size of the test dataset}. The prediction results are shown in Fig.\ref{fig:result_CGLMP}, in which the blue line and the {orange line} represent the prediction and the exact values of coherent information, respectively. It can be seen that our model behaves very well in this task, whose MSE is 0.0040.

As a comparison, the above setting can also be handled by the semi-device-independent approach given by Ref.\cite{wei2021analytic}. As illustrated by the green line in Fig.\ref{fig:result_CGLMP}, the method in Ref.\cite{wei2021analytic} can provide nontrivial lower bound for the coherent information only when $\epsilon$ ranges from {0.8610} to 1, and the performance is also much worse. Therefore, in this case the neural network approach beats the approach in Ref.\cite{wei2021analytic} easily.

\begin{figure}[!ht]
	\centering
	\includegraphics[width=0.45\textwidth]{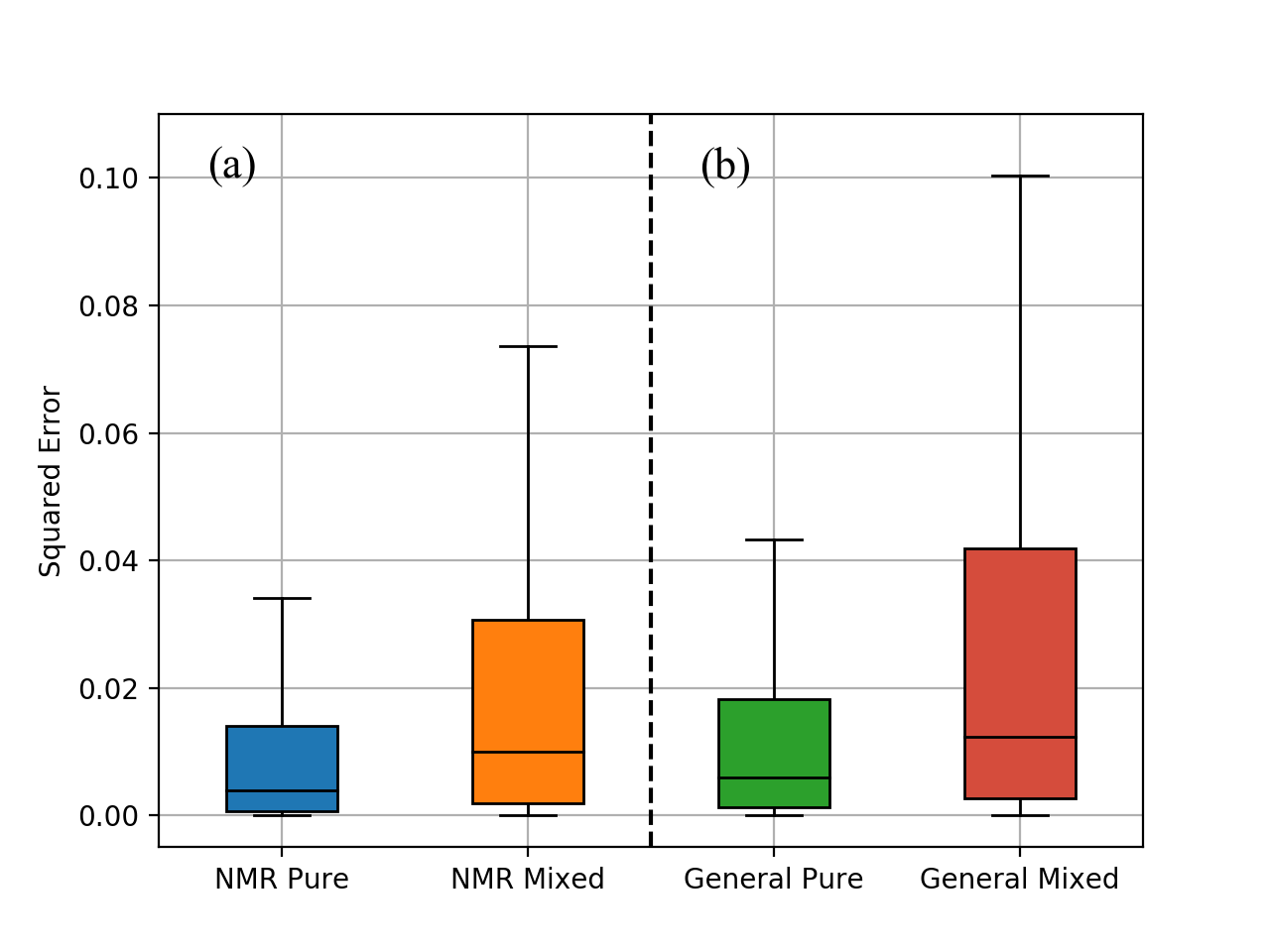}
	\caption{The neural network predictions for the coherent information of qutrit-qutrit quantum states. Each boxplot indicates the degree of dispersion and skewness in the squared errors for each case. (a) Boxplots of NMR pure states (blue) and NMR mixed states (orange), where the corresponding MSEs are {0.0137 and 0.0281} respectively. (b) Boxplots of general pure states (green) and general mixed states (red), where the corresponding MSEs are {0.0161 and 0.0407} respectively.}
	\label{fig:result_Qutrit}
\end{figure}

We now apply neural networks {to} quantify the coherent information for more general qutrit-qutrit quantum states. For this, we consider the following four different classes of quantum states.
\begin{itemize}
\item $(1-\epsilon)I_9/9+\epsilon|\psi\rangle\langle\psi|$, where $|\psi\rangle$ is an arbitrary pure state (called \emph{NMR pure states}).

\item $(1-\epsilon)I_9/9+\epsilon\rho_0$, where $\rho_0$ is an arbitrary mixed state (called \emph{NMR mixed states}).

\item $|\psi\rangle\langle\psi|$, where $|\psi\rangle$ is an arbitrary pure state (called \emph{general pure states}).

\item $\rho_0$, where $\rho_0$ is an arbitrary mixed states (called \emph{general mixed states}).
\end{itemize}

In order to make the training data as representative as possible, we will generate them by picking up sample states from all the four classes. After the training stage is finished, we then test the performance of our neural network for each class separately. The neural network we choose has the same structure as before, whose 4 hidden layers has 400, 200, 100, 50 neurons, respectively.

More concretely, when generating training data from the classes of NMR pure and NMR mixed states, the way we pick up sample states is similar to the previous experiment. That is, we select $\epsilon$ at intervals 0.025 and randomly generate 2,000 states for each $\epsilon$ from 0.2 to 0.75, and 4,000 states for each $\epsilon$ from 0.775 to 0.975, where 82,000 sample states are picked up for each class. When generating training data from the classes of general pure and general mixed states, whose coherent information considered for training {dataset} ranges from 0 to $\log_23$ and from $-\log_23$ to 1.5 respectively, we sample states according to coherent information, and choose 4,000 states randomly in each interval of size 0.1. Here the only exception is for {general} mixed states with coherent information larger than 1.5. Due to the low sampling efficiency, we do not insist on the number {of states} produced in this case, {and just collect all such states during the sampling process.} Eventually, 64,000 general pure states, 124,211 general mixed states, and 164,000 NMR pure and mixed states are chosen as training states. Performing the same local measurements as the previous experiment on them and recording the outcome statistics data for each state as the {features}, together with the corresponding coherent information as the label, we obtain the whole training dataset, {which has 352,211 sample states.}

After training the model, we now test its performance on the above four classes of quantum states separately. That is to say, for each class we generate a test {dataset} by sampling {around} 2,000 random states uniformly according to coherent information. Similarly, we use the MSE to measure the prediction performance of our model. The results are listed in Fig.\ref{fig:result_Qutrit}. For convenience, we represent the experimental results in the form of boxplot, where each boxplot displays a dataset based on a five-number summary: the lowest data point excluding any outliers (minimum), the median of the lower half of the dataset (first quartile), the median of the dataset, the median of the upper half of the dataset (third quartile) and the largest data point excluding any outliers (maximum).

As we can see in Fig.\ref{fig:result_Qutrit}(a), when predicting the coherent information of NMR pure and mixed states, the MSEs are {0.0137 and 0.0281} respectively. Here the distribution of the squared error for NMR pure states is very narrow, indicating the very high quality predictions. Fig.\ref{fig:result_Qutrit}(b) shows the results for general pure states and general mixed states, where the corresponding MSEs are {0.0161 and 0.0407 respectively}. Overall, it can be said that even for the most general case of qutrit-qutrit quantum states, the neural network model still enjoys decent performance in predicting coherent information.

\subsection{Higher-dimensional cases}

\begin{table*}[htb] \scriptsize
    \centering
	\caption {The structure and configuration details of the convolutional neural network (see Refs. \cite{lecun1998gradient,albawi2017understanding} for introductions to convolutional neural networks).}
	\setlength{\tabcolsep}{4.5mm}
   {
	\begin{tabular}{lccccccl}
		\hline
		\hline
		Layers &Type&Neurons&Filters&Kernel size&Strides&Pool size\\
		\hline
		0-1 & Convolution2D & (None, 9, 9, 32) & 32  & 2$\times 2$ & $1\times 1$ & -\\
		1-2 & Max-pooling2D & (None, 8, 8, 32) & - & - & $1\times 1$ & 2$\times 2$\\
		2-3 & Convolution2D & (None, 7, 7, 64) & 64  & 2$\times 2$ & $1\times 1$ & -\\
		3-4 & Max-pooling2D & (None, 6, 6, 64) & - & - & $1\times 1$ & 2$\times 2$\\
		4-5 & Convolution2D & (None, 5, 5, 64)  & 64  & 2$\times 2$ & $1\times 1$ & -\\
		5-6 & Fully-connected & (None, 32) & -  & - & - & -\\
		6-7 & Fully-connected & (None, 1) & -  & - & - & -\\
		\hline
		\hline
	\end{tabular}}
	\label{table:CNN}
	\end{table*}
	
We now utilize neural networks to predict the coherent information of bipartite quantum states in higher-dimensional Hilbert space $\mathcal{H}^d\otimes\mathcal{H}^d$. In this case, the task of picking up proper training dataset is more challenging. In order to make the sample states for training as representative as possible, we consider a class of quantum states of the form
\begin{eqnarray}\label{training_state}
\rho_d=\alpha\rho_0+\beta I_{d^2}/{d^2}+\gamma|\psi^{me}_d\rangle\langle\psi^{me}_d|,
\end{eqnarray}
where $\alpha,\beta,\gamma \geq 0$, $\alpha+\beta+\gamma =1$, $\rho_0$ is an arbitrary quantum state {in $\mathcal{H}^d\otimes\mathcal{H}^d$}, $I_{d^2}/{d^2}$ is the bipartite maximally mixed state, and $|\psi^{me}_d\rangle=\sum_{i=0}^{d-1}|ii\rangle /\sqrt{d}$ is the maximally entangled state. In other words, $\rho_d$ is a linear combination of three completely different types of quantum states.

To demonstrate how to generate training dataset for this case, let us take $d=5$ as an example. The training states will be composed by two parts. Firstly, we generate the first part of the training states by sampling {$\rho_5$} in Eq.\eqref{training_state}, where about 170,000 states are chosen. More concretely, when $\alpha$ is no less than 0.4, $\alpha$ and $\beta$ are varied at intervals 0.02, and for each choices of the coefficients, 200 random $\rho_5$ are produced by sampling $\rho_0$. When $\alpha$ is smaller than 0.4, we vary $\alpha,\beta$ at intervals 0.04 and sample 100 $\rho_5$ states similarly for each choices of the coefficients. The second part of the training states will be produced in a similar way with the qutrit-qutrit case. That is, we separately sample 82,000 NMR pure states, 130,000 NMR mixed states, 40,000 general pure states, and 84,000 general mixed states {in such a way that their coherent information is distributed evenly}. Here, when sampling general pure states, we find that it is hard to sample states with coherent information less than 0.8. In order to have a relatively complete coherent information distribution, we generate general pure states by
\begin{eqnarray}\label{pure_state}
|\psi_5\rangle=\alpha|\psi_5'\rangle + (1-\alpha)|\psi_5^{sep}\rangle,
\end{eqnarray}
where $\alpha\in[0,1]$, $|\psi_5'\rangle$ and $|\psi_5^{sep}\rangle$ are an arbitrary pure state and an arbitrary separable pure state in $\mathcal{H}^5\otimes\mathcal{H}^5$, respectively. After picking up the {states for training}, we measure them by the local measurements given in {Eqs.\eqref{CGLMP_Measurement1} and \eqref{CGLMP_Measurement2}}, and generate the measurement outcome statistics data as the features for the neural network. As training dataset, we compute the coherent information for each sample state as the corresponding label.

Compared with fully connected neural networks, in higher-dimensional cases we find that convolutional neural networks have better performance in quantifying entanglement (see Refs. \cite{lecun1998gradient,albawi2017understanding} for introductions to convolutional neural networks). The structure of the convolutional neural network we choose is listed in Table \ref{table:CNN}. Again, we test the performance of the trained neural network by utilizing it to separately predict the coherent information of four classes of quantum states, say NMR pure states, NMR mixed states, general pure states and general mixed states.

\begin{figure}[!ht]
	\centering
	\includegraphics[width=0.45\textwidth]{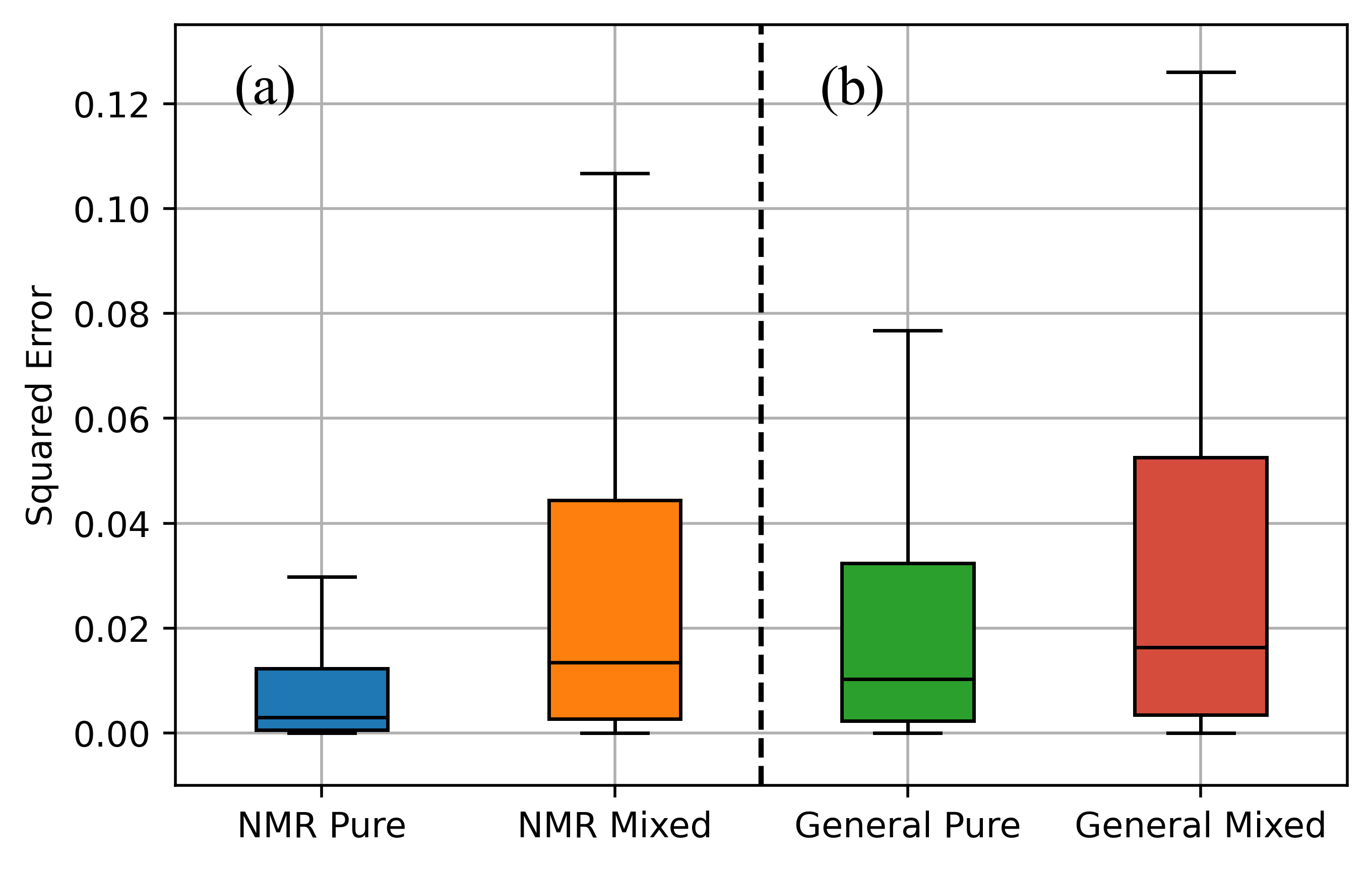}
	\caption{The neural network predictions for the coherent information of higher-dimensional quantum states, where $d=5$. Each boxplot indicates the degree of dispersion and skewness in the squared errors for each case. (a) Boxplots of NMR pure states (blue) and NMR
    mixed states (orange), where the corresponding MSEs are {0.0123 and 0.0356} respectively. (b) Boxplots of general pure states (green) and general mixed states (red), where the corresponding MSEs are {0.0275 and 0.0436} respectively.}
	\label{fig:result_qudit_boxplot_CNN}
\end{figure}

For each class of test states, we fix the number of instances to be 2,000, and the experimental results {are} presented in Fig.\ref{fig:result_qudit_boxplot_CNN}. It can be seen that the overall performance is similar {to} the qutrit-qutrit case, and the MSEs are {0.0123, 0.0356, 0.0275 and 0.0436}, respectively.

In addition to the boxplots, to further describe the degree of dispersion of the prediction results, we also compute the relative error of predictions for test states. For the class of general mixed states, which is the hardest to predict, the mean relative error is smaller than {$15\%$ $(10\%)$ for test states whose absolute values of the exact coherent information are larger than 0.5 (1.0).}

To examine the performance of neural networks better, we now increase the {dimension} of target quantum states even further, and use a convolutional neural network similar {to} the one in Table \ref{table:CNN} to predict coherent information, where the only difference is that we increase the kernel size and pool size from $2\times2$ to $3\times3$. Note that when quantum dimension goes up, it is very challenging to handle the sharply-increasing amount of data. Therefore, from now on we only do the following two experiments: (1) both the training states and the test states are sampled from general mixed states; (2) both the training states and the test states are sampled from general pure states. In each case, the number of training samples is around 40,000, and the number of test samples is {over} 2,000. Similar {to} before, we try to pick up these samples evenly according to their coherent information.

We run the model for quantum states with dimensions from 5 to 10, and the results are listed in Table \ref{table:result_diffdim}. Compared with Fig.\ref{fig:result_qudit_boxplot_CNN}(b), the {prediction MSEs for general pure states and general mixed states are improved from 0.0275 to 0.0135 and from 0.0436 to 0.0347 respectively, showing that} utilizing prior knowledge on target quantum states can improve the performance. Furthermore, a very surprising fact is that, unlike fully connected neural networks, when the quantum dimension goes up, the convolutional neural network model behaves even better, confirming the claim that convolutional neural network is a more suitable choice for higher-dimensional cases. Combining all the results together, we can see that neural networks still enjoy a very decent performance in quantifying entanglement even for high-dimensional quantum systems.

\begin{table}[h] \scriptsize
\caption{The MSEs of predicting coherent information for different dimensions. Each case is trained with around 40,000 samples and tested with over 2,000 samples.}
\setlength{\tabcolsep}{1.8mm}{
\begin{tabular}{lcccccc}
	\hline
	\hline
	Dimension & $d=5$ & $d=6$ & $d=7$ & $d=8$ & $d=9$ & $d=10$\\
	\hline
	Pure & 0.0135 & 0.0110 & 0.0107  & 0.0090 & 0.0093 & 0.0086\\
	Mixed & 0.0347 & 0.0321 & 0.0255 & 0.0217 & 0.0171 & 0.0149\\
	\hline
	\hline
\end{tabular}}
\label{table:result_diffdim}
\end{table}

\section{Quantifying Multipartite Entanglement}

Compared with bipartite entanglement, multipartite entanglement has much more complicated mathematical structures. Many different measures have been defined to quantify multipartite entanglement, and almost all of them are hard to calculate. As mentioned before, in order to apply neural network models {to quantify} unknown multipartite entanglement, we have to make sure that correct labels, which is some entanglement measure here, of training states {are computationally tractable}. For this, we choose the geometric measure of entanglement as our target entanglement measure, as it is relatively easy to handle among well-known multipartite entanglement measures, especially for the case of pure states. Therefore, in this section we will only train and test our neural networks using three-qubit pure states.

To prepare training data, for every three-qubit pure state $|\psi\rangle\in\mathcal{H}^2\otimes\mathcal{H}^2\otimes\mathcal{H}^2$ we measure each qubit by observables Pauli $\sigma_x$ and $\sigma_y$, and record the outcome statistics data as the corresponding {features}. As the label {of} the above {features}, we compute the value of $E_G(|\psi\rangle)$ in Eq.(\ref{GME_Pure}) by numerical calculations. The whole training dataset is {generated by} sampling states according to their GME. Concretely, when the value of GME is from 0 to 0.7, we sample 5,000 three-qubit pure states for each interval of size 0.1. When the GME is larger than 0.7, we do not insist on the number of samples obtained, and just collect all the qualified states we sample. In addition, we also sample 5,000 separable states as part of the training states, and the involved local measurements are the same {as} before. Combining the above two parts, we get the training dataset with 40,342 instances. Then we use the training data to train a 4-hidden-layer {fully connected} neural network, where each layer has 50, 20, 10, 5 neurons, respectively.

We now turn to the sampling of test states. Recall that there are only two classes of genuinely tripartite entanglement which are inequivalent under local operations and classical communication(LOCC) \cite{dur2000three}. One class is represented by the Greenberger-Horne-Zeilinger (GHZ) state \cite{greenberger1989going},
\begin{eqnarray}\label{GHZ}
|{\rm GHZ}\rangle=\left(|000\rangle+|111\rangle\right)/\sqrt{2},
\end{eqnarray}
and the other class is represented by the W state \cite{dur2000three},
\begin{eqnarray}\label{W}
|{\rm W}\rangle=\left(|001\rangle+|010\rangle+|100\rangle\right)/\sqrt{3}.
\end{eqnarray}
Therefore, we are interested in the {behavior} of our model on these two states. For this, we consider the following two classes of states,
\begin{eqnarray}\label{W_W_bar}
|\varphi_p\rangle=\sqrt{p}|{\rm W}\rangle+\sqrt{1-p}|{\rm \overline{W}}\rangle,
\end{eqnarray}
and
\begin{multline}\label{psi_3}
|\varphi'_p\rangle=\sqrt{p}|{\rm GHZ}\rangle+\sqrt{(1-p)/2}\left(|{\rm W}\rangle+|{\rm \overline{W}}\rangle\right),
\end{multline}
where $p\in[0,1]$ and $|{\rm \overline{W}}\rangle=\left(|110\rangle+|101\rangle+|011\rangle\right)/\sqrt{3}$.

\begin{figure}[!ht]
	\centering
	\includegraphics[width=0.45\textwidth]{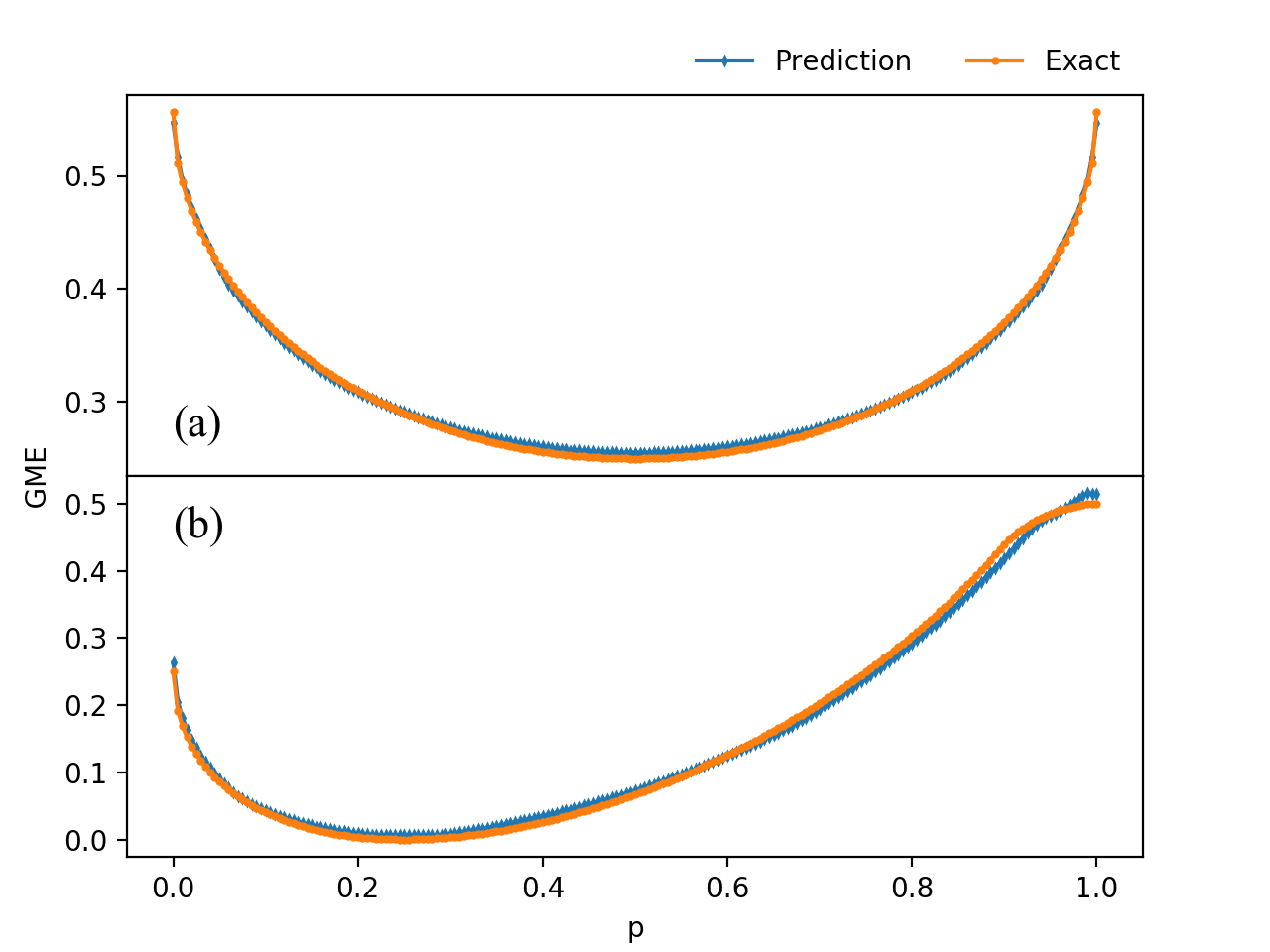}
	\caption{The neural network predictions for the GME of three-qubit pure states. The blue line and the orange line represent the prediction values and the exact values, respectively. (a) The results for the state $|\varphi_p\rangle$, where the MSE of the prediction values is $1.26\times 10^{-5}$. (b) The results for the state $|\varphi'_p\rangle$, where the MSE of the prediction values is $7.72\times 10^{-5}$.}
	\label{fig:result_GME}
\end{figure}
We sample states within $|\varphi_p\rangle$ and $|\varphi'_p\rangle$ to serve as our test samples, where $p$ is chosen from 0 to 1 at intervals 0.005. The prediction results are illustrated in Fig.\ref{fig:result_GME}, in which (a) is for the state $|\varphi_p\rangle$ and (b) is for the state $|\varphi'_p\rangle$. As we can see, the prediction values fit the exact values very well in both cases, whose MSEs are $1.26\times 10^{-5}$ and $7.72\times 10^{-5}$, respectively.

In addition, we also test the performance of a properly trained neural network on another test {dataset} composed by around 2,000 random three-qubit pure states, whose GME ranges from 0 to 0.7 evenly. Again, the performance is excellent and the MSE is only 0.0043. These high quality predictions strongly imply that neural networks can have {satisfactory} performance in quantifying unknown multipartite entanglement.

\section{Neural Networks can feel Nonlocality}

In our neural network models, the features are chosen as statistics data of the outcomes when local measurements are performed on samples. Actually this kind of statistics data is exactly what people utilize in studying nonlocality, though as stressed before, quantum nonlocality is not considered in the training and testing of our neural networks at all. However, due to this similarity, we may wonder, does quantum nonlocality play any role in the workings of our neural networks? Interestingly, we observe strong evidences suggesting that the answer to this question is positive.

For this, let us go back to the neural network we have trained {for state $\rho^{mv}_{3_{\epsilon}}$} in Sec III. A. We now run that well-trained neural network on the following special class of quantum states,
\begin{eqnarray}\label{CGLMP_Class_2}
\rho_{{p,\gamma}}=(1-p)I_9/9+p|\phi_{{\gamma}}\rangle\langle\phi_{{\gamma}}|,
\end{eqnarray}
where
\begin{eqnarray}\label{CGLMP_State_2}
|\phi_{{\gamma}}\rangle=\gamma(|00\rangle+|11\rangle)+\sqrt{1-2\gamma^2}|22\rangle,
\end{eqnarray}
$\gamma\in[0.6, \sqrt{2}/2]$ and $p$ is chosen such that the coherent information of $\rho_{{p,\gamma}}$ is positive. Again, for any test state in the $\rho_{{p,\gamma}}$ class, we measure it with obsevables $A_a$ and $B_b$ given by {Eqs.\eqref{CGLMP_Measurement1} and \eqref{CGLMP_Measurement2}} to generate the features for the neural network. Meanwhile, the violation of this test state to the CGLMP inequality is also computed, which as usual is used to measure quantum nonlocality. Then, after running the neural network on these test samples, we have the chance to examine whether the performance of prediction has {anything} to do with quantum nonlocality.

\begin{figure}[!ht]
	\centering
	\includegraphics[width=0.45\textwidth]{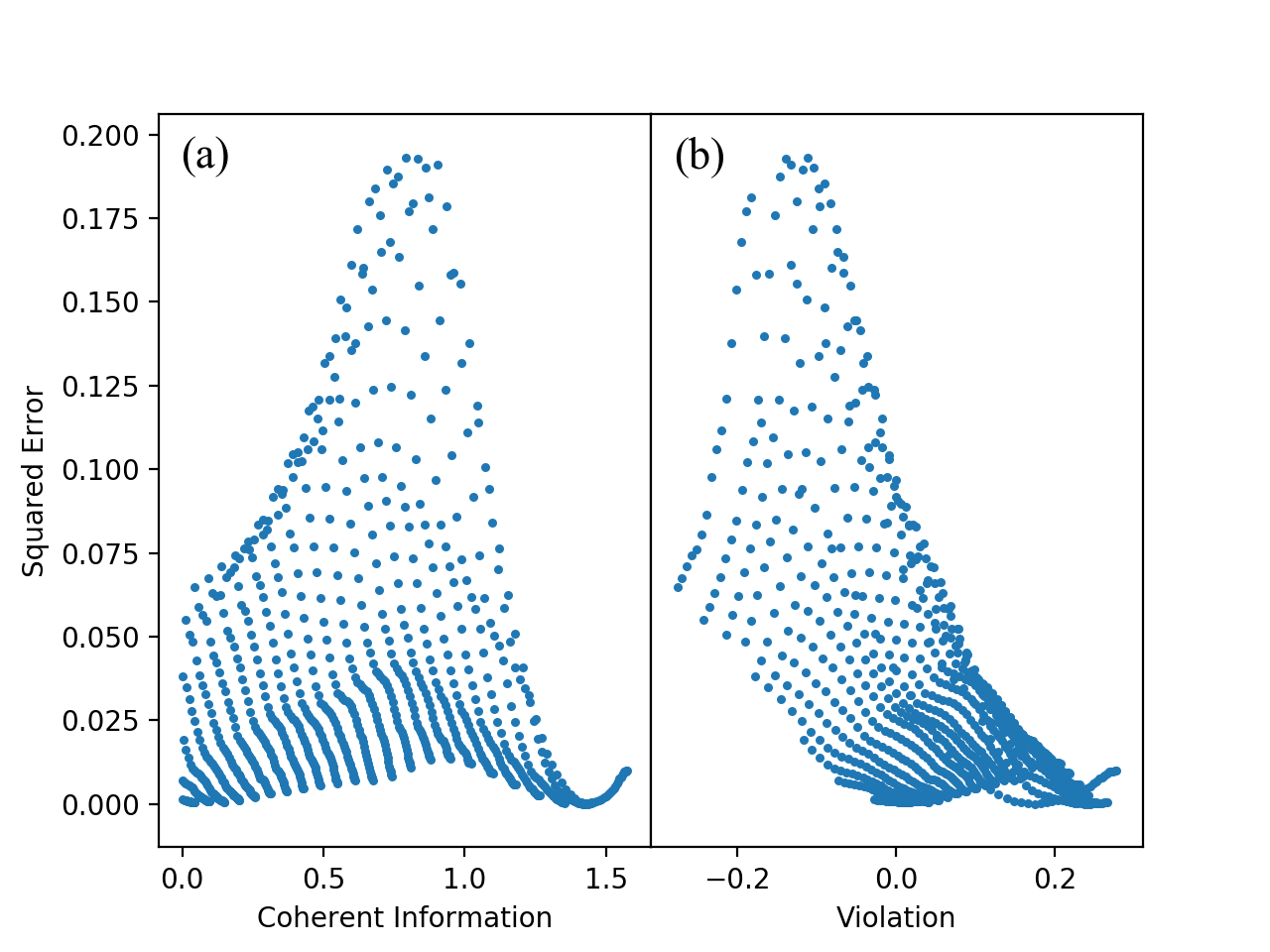}
	\caption{The relations between coherent information, nonlocality, and squared error. Each blue dot represents a fixed parameter pair $(p,\gamma)$ for $\rho_{{p,\gamma}}$. (a) The relation between squared error and coherent information, where the PCC is 0.0010. (b) The relation between squared error and violation, where the PCC is -0.6041.}
	\label{fig:result_nonlocality}
\end{figure}

The results are listed in Fig.\ref{fig:result_nonlocality}. Here every pair of $(p,\gamma)$, which corresponds to a test quantum state, is represented by a blue dot. Along the $x$-axis, in Fig.\ref{fig:result_nonlocality}(a) we record the coherent information, and in Fig.\ref{fig:result_nonlocality}(b) we record the violation. It can be seen clearly that when a test state has stronger quantum nonlocality, the neural network tends to predict its coherent information more accurately.

To further {analyze} the statistical connections between nonlocality, coherent information, and squared error, we utilize the Pearson correlation coefficient (PCC) to analyze the data, which is a measure of linear correlation between the paired data $\{(x_1,y_1),...,(x_n,y_n)\}$ defined as
\begin{eqnarray}\label{PCC}
r_{xy}=\frac{\sum_{i=1}^n(x_i-\bar{x})(y_i-\bar{y})}{\sqrt{\sum_{i=1}^n(x_i-\bar{x})^2}\sqrt{\sum_{i=1}^n(y_i-\bar{y})^2}},
\end{eqnarray}
where $\bar{x}=\frac{1}{n}\sum_{i=1}^nx_i$ and analogously for $\bar{y}$. We find that the PCC between squared error and coherent information is 0.0010, implying that almost no linear correlation exists between them. As a sharp difference, the PCC between squared error and violation is -0.6041, confirming that a strong correlation exists between them.

We stress again that when training the neural networks, we do not provide any knowledge {of} quantum nonlocality. However, the above evidences show that in our neural networks, stronger nonlocality in test samples tends to result in better prediction quality, which is quite surprising.

\section{Discussions}

In this paper, we have showed strong evidences implying that neural networks can be used to quantify unknown quantum entanglement after {being} properly trained, where the performance is very decent, and can be much better than that of some previous approaches like semi-device-independent protocols.

It should be pointed out that, when training neural networks, we only need to measure sample quantum states locally and then collect the {outcome} statistics data as the corresponding features, which is actually quite experiment-friendly, as we do not have to take care of the experimental loophole issues like in Bell experiments studying quantum nonlocality~\cite{hensen2015loophole,giustina2015significant,shalm2015strong}. Furthermore, to obtain the data needed by neural networks with sufficient precision, the required number of measurements is reasonable. Therefore, with the rise of quantum engineering in recent years, it is very realistic to consider applying neural networks widely in quantum engineering problems where a rough 
quantification of underlying entanglement is helpful. Because of this, it will be interesting to improve the performance of current neural networks further, and try to utilize neural networks to quantify more measures of quantum entanglement.

\begin{acknowledgements}
This work was supported by the National Key R\&D Program of China, Grant No. 2018YFA0306703, the National Natural Science Foundation of China, Grant No. 20181311604, and the start-up funds of Tsinghua University, Grant No. 53330100118. This work was also supported in part by the Zhongguancun Haihua Institute for Frontier Information Technology.
\end{acknowledgements}

\bibliography{Quantifying_unknown_entanglement_by_neural_networks}

\begin{thebibliography}{45}%
\makeatletter
\providecommand \@ifxundefined [1]{%
 \@ifx{#1\undefined}
}%
\providecommand \@ifnum [1]{%
 \ifnum #1\expandafter \@firstoftwo
 \else \expandafter \@secondoftwo
 \fi
}%
\providecommand \@ifx [1]{%
 \ifx #1\expandafter \@firstoftwo
 \else \expandafter \@secondoftwo
 \fi
}%
\providecommand \natexlab [1]{#1}%
\providecommand \enquote  [1]{``#1''}%
\providecommand \bibnamefont  [1]{#1}%
\providecommand \bibfnamefont [1]{#1}%
\providecommand \citenamefont [1]{#1}%
\providecommand \href@noop [0]{\@secondoftwo}%
\providecommand \href [0]{\begingroup \@sanitize@url \@href}%
\providecommand \@href[1]{\@@startlink{#1}\@@href}%
\providecommand \@@href[1]{\endgroup#1\@@endlink}%
\providecommand \@sanitize@url [0]{\catcode `\\12\catcode `\$12\catcode
  `\&12\catcode `\#12\catcode `\^12\catcode `\_12\catcode `\%12\relax}%
\providecommand \@@startlink[1]{}%
\providecommand \@@endlink[0]{}%
\providecommand \url  [0]{\begingroup\@sanitize@url \@url }%
\providecommand \@url [1]{\endgroup\@href {#1}{\urlprefix }}%
\providecommand \urlprefix  [0]{URL }%
\providecommand \Eprint [0]{\href }%
\providecommand \doibase [0]{https://doi.org/}%
\providecommand \selectlanguage [0]{\@gobble}%
\providecommand \bibinfo  [0]{\@secondoftwo}%
\providecommand \bibfield  [0]{\@secondoftwo}%
\providecommand \translation [1]{[#1]}%
\providecommand \BibitemOpen [0]{}%
\providecommand \bibitemStop [0]{}%
\providecommand \bibitemNoStop [0]{.\EOS\space}%
\providecommand \EOS [0]{\spacefactor3000\relax}%
\providecommand \BibitemShut  [1]{\csname bibitem#1\endcsname}%
\let\auto@bib@innerbib\@empty
\bibitem [{\citenamefont {Chuang}\ and\ \citenamefont
  {Nielsen}(1997)}]{chuang1997prescription}%
  \BibitemOpen
  \bibfield  {author} {\bibinfo {author} {\bibfnamefont {I.~L.}\ \bibnamefont
  {Chuang}}\ and\ \bibinfo {author} {\bibfnamefont {M.~A.}\ \bibnamefont
  {Nielsen}},\ }\bibfield  {title} {\bibinfo {title} {Prescription for
  experimental determination of the dynamics of a quantum black box},\
  }\href@noop {} {\bibfield  {journal} {\bibinfo  {journal} {Journal of Modern
  Optics}\ }\textbf {\bibinfo {volume} {44}},\ \bibinfo {pages} {2455}
  (\bibinfo {year} {1997})}\BibitemShut {NoStop}%
\bibitem [{\citenamefont {Poyatos}\ \emph {et~al.}(1997)\citenamefont
  {Poyatos}, \citenamefont {Cirac},\ and\ \citenamefont
  {Zoller}}]{poyatos1997complete}%
  \BibitemOpen
  \bibfield  {author} {\bibinfo {author} {\bibfnamefont {J.}~\bibnamefont
  {Poyatos}}, \bibinfo {author} {\bibfnamefont {J.~I.}\ \bibnamefont {Cirac}},\
  and\ \bibinfo {author} {\bibfnamefont {P.}~\bibnamefont {Zoller}},\
  }\bibfield  {title} {\bibinfo {title} {Complete characterization of a quantum
  process: the two-bit quantum gate},\ }\href@noop {} {\bibfield  {journal}
  {\bibinfo  {journal} {Physical Review Letters}\ }\textbf {\bibinfo {volume}
  {78}},\ \bibinfo {pages} {390} (\bibinfo {year} {1997})}\BibitemShut
  {NoStop}%
\bibitem [{\citenamefont {G{\"u}hne}\ and\ \citenamefont
  {T{\'o}th}(2009)}]{guhne2009entanglement}%
  \BibitemOpen
  \bibfield  {author} {\bibinfo {author} {\bibfnamefont {O.}~\bibnamefont
  {G{\"u}hne}}\ and\ \bibinfo {author} {\bibfnamefont {G.}~\bibnamefont
  {T{\'o}th}},\ }\bibfield  {title} {\bibinfo {title} {Entanglement
  detection},\ }\href@noop {} {\bibfield  {journal} {\bibinfo  {journal}
  {Physics Reports}\ }\textbf {\bibinfo {volume} {474}},\ \bibinfo {pages} {1}
  (\bibinfo {year} {2009})}\BibitemShut {NoStop}%
\bibitem [{\citenamefont {Rosset}\ \emph {et~al.}(2012)\citenamefont {Rosset},
  \citenamefont {Ferretti-Sch{\"o}bitz}, \citenamefont {Bancal}, \citenamefont
  {Gisin},\ and\ \citenamefont {Liang}}]{rosset2012imperfect}%
  \BibitemOpen
  \bibfield  {author} {\bibinfo {author} {\bibfnamefont {D.}~\bibnamefont
  {Rosset}}, \bibinfo {author} {\bibfnamefont {R.}~\bibnamefont
  {Ferretti-Sch{\"o}bitz}}, \bibinfo {author} {\bibfnamefont {J.-D.}\
  \bibnamefont {Bancal}}, \bibinfo {author} {\bibfnamefont {N.}~\bibnamefont
  {Gisin}},\ and\ \bibinfo {author} {\bibfnamefont {Y.-C.}\ \bibnamefont
  {Liang}},\ }\bibfield  {title} {\bibinfo {title} {Imperfect measurement
  settings: Implications for quantum state tomography and entanglement
  witnesses},\ }\href@noop {} {\bibfield  {journal} {\bibinfo  {journal}
  {Physical Review A}\ }\textbf {\bibinfo {volume} {86}},\ \bibinfo {pages}
  {062325} (\bibinfo {year} {2012})}\BibitemShut {NoStop}%
\bibitem [{\citenamefont {Collins}\ \emph
  {et~al.}(2002{\natexlab{a}})\citenamefont {Collins}, \citenamefont {Gisin},
  \citenamefont {Popescu}, \citenamefont {Roberts},\ and\ \citenamefont
  {Scarani}}]{collins2002nonsep}%
  \BibitemOpen
  \bibfield  {author} {\bibinfo {author} {\bibfnamefont {D.}~\bibnamefont
  {Collins}}, \bibinfo {author} {\bibfnamefont {N.}~\bibnamefont {Gisin}},
  \bibinfo {author} {\bibfnamefont {S.}~\bibnamefont {Popescu}}, \bibinfo
  {author} {\bibfnamefont {D.}~\bibnamefont {Roberts}},\ and\ \bibinfo {author}
  {\bibfnamefont {V.}~\bibnamefont {Scarani}},\ }\bibfield  {title} {\bibinfo
  {title} {Bell-type inequalities to detect true n-body nonseparability},\
  }\href@noop {} {\bibfield  {journal} {\bibinfo  {journal} {Physical Review
  Letters}\ }\textbf {\bibinfo {volume} {88}},\ \bibinfo {pages} {170405}
  (\bibinfo {year} {2002}{\natexlab{a}})}\BibitemShut {NoStop}%
\bibitem [{\citenamefont {Bancal}\ \emph {et~al.}(2011)\citenamefont {Bancal},
  \citenamefont {Gisin}, \citenamefont {Liang},\ and\ \citenamefont
  {Pironio}}]{bancal2011device}%
  \BibitemOpen
  \bibfield  {author} {\bibinfo {author} {\bibfnamefont {J.-D.}\ \bibnamefont
  {Bancal}}, \bibinfo {author} {\bibfnamefont {N.}~\bibnamefont {Gisin}},
  \bibinfo {author} {\bibfnamefont {Y.-C.}\ \bibnamefont {Liang}},\ and\
  \bibinfo {author} {\bibfnamefont {S.}~\bibnamefont {Pironio}},\ }\bibfield
  {title} {\bibinfo {title} {Device-independent witnesses of genuine
  multipartite entanglement},\ }\href@noop {} {\bibfield  {journal} {\bibinfo
  {journal} {Physical Review Letters}\ }\textbf {\bibinfo {volume} {106}},\
  \bibinfo {pages} {250404} (\bibinfo {year} {2011})}\BibitemShut {NoStop}%
\bibitem [{\citenamefont {P{\'a}l}\ and\ \citenamefont
  {V{\'e}rtesi}(2011)}]{pal2011multisetting}%
  \BibitemOpen
  \bibfield  {author} {\bibinfo {author} {\bibfnamefont {K.~F.}\ \bibnamefont
  {P{\'a}l}}\ and\ \bibinfo {author} {\bibfnamefont {T.}~\bibnamefont
  {V{\'e}rtesi}},\ }\bibfield  {title} {\bibinfo {title} {Multisetting
  bell-type inequalities for detecting genuine multipartite entanglement},\
  }\href@noop {} {\bibfield  {journal} {\bibinfo  {journal} {Physical Review
  A}\ }\textbf {\bibinfo {volume} {83}},\ \bibinfo {pages} {062123} (\bibinfo
  {year} {2011})}\BibitemShut {NoStop}%
\bibitem [{\citenamefont {Murta}\ \emph {et~al.}(2016)\citenamefont {Murta},
  \citenamefont {Ramanathan}, \citenamefont {M{\'o}ller},\ and\ \citenamefont
  {Cunha}}]{murta2016quantum}%
  \BibitemOpen
  \bibfield  {author} {\bibinfo {author} {\bibfnamefont {G.}~\bibnamefont
  {Murta}}, \bibinfo {author} {\bibfnamefont {R.}~\bibnamefont {Ramanathan}},
  \bibinfo {author} {\bibfnamefont {N.}~\bibnamefont {M{\'o}ller}},\ and\
  \bibinfo {author} {\bibfnamefont {M.~T.}\ \bibnamefont {Cunha}},\ }\bibfield
  {title} {\bibinfo {title} {Quantum bounds on multiplayer linear games and
  device-independent witness of genuine tripartite entanglement},\ }\href@noop
  {} {\bibfield  {journal} {\bibinfo  {journal} {Physical Review A}\ }\textbf
  {\bibinfo {volume} {93}},\ \bibinfo {pages} {022305} (\bibinfo {year}
  {2016})}\BibitemShut {NoStop}%
\bibitem [{\citenamefont {Baccari}\ \emph {et~al.}(2017)\citenamefont
  {Baccari}, \citenamefont {Cavalcanti}, \citenamefont {Wittek},\ and\
  \citenamefont {Ac{\'\i}n}}]{baccari2017efficient}%
  \BibitemOpen
  \bibfield  {author} {\bibinfo {author} {\bibfnamefont {F.}~\bibnamefont
  {Baccari}}, \bibinfo {author} {\bibfnamefont {D.}~\bibnamefont {Cavalcanti}},
  \bibinfo {author} {\bibfnamefont {P.}~\bibnamefont {Wittek}},\ and\ \bibinfo
  {author} {\bibfnamefont {A.}~\bibnamefont {Ac{\'\i}n}},\ }\bibfield  {title}
  {\bibinfo {title} {Efficient device-independent entanglement detection for
  multipartite systems},\ }\href@noop {} {\bibfield  {journal} {\bibinfo
  {journal} {Physical Review X}\ }\textbf {\bibinfo {volume} {7}},\ \bibinfo
  {pages} {021042} (\bibinfo {year} {2017})}\BibitemShut {NoStop}%
\bibitem [{\citenamefont {Tavakoli}\ \emph {et~al.}(2018)\citenamefont
  {Tavakoli}, \citenamefont {Abbott}, \citenamefont {Renou}, \citenamefont
  {Gisin},\ and\ \citenamefont {Brunner}}]{tavakoli2018semi}%
  \BibitemOpen
  \bibfield  {author} {\bibinfo {author} {\bibfnamefont {A.}~\bibnamefont
  {Tavakoli}}, \bibinfo {author} {\bibfnamefont {A.~A.}\ \bibnamefont
  {Abbott}}, \bibinfo {author} {\bibfnamefont {M.-O.}\ \bibnamefont {Renou}},
  \bibinfo {author} {\bibfnamefont {N.}~\bibnamefont {Gisin}},\ and\ \bibinfo
  {author} {\bibfnamefont {N.}~\bibnamefont {Brunner}},\ }\bibfield  {title}
  {\bibinfo {title} {Semi-device-independent characterization of multipartite
  entanglement of states and measurements},\ }\href@noop {} {\bibfield
  {journal} {\bibinfo  {journal} {Physical Review A}\ }\textbf {\bibinfo
  {volume} {98}},\ \bibinfo {pages} {052333} (\bibinfo {year}
  {2018})}\BibitemShut {NoStop}%
\bibitem [{\citenamefont {Zwerger}\ \emph {et~al.}(2019)\citenamefont
  {Zwerger}, \citenamefont {D{\"u}r}, \citenamefont {Bancal},\ and\
  \citenamefont {Sekatski}}]{zwerger2019device}%
  \BibitemOpen
  \bibfield  {author} {\bibinfo {author} {\bibfnamefont {M.}~\bibnamefont
  {Zwerger}}, \bibinfo {author} {\bibfnamefont {W.}~\bibnamefont {D{\"u}r}},
  \bibinfo {author} {\bibfnamefont {J.-D.}\ \bibnamefont {Bancal}},\ and\
  \bibinfo {author} {\bibfnamefont {P.}~\bibnamefont {Sekatski}},\ }\bibfield
  {title} {\bibinfo {title} {Device-independent detection of genuine
  multipartite entanglement for all pure states},\ }\href@noop {} {\bibfield
  {journal} {\bibinfo  {journal} {Physical Review Letters}\ }\textbf {\bibinfo
  {volume} {122}},\ \bibinfo {pages} {060502} (\bibinfo {year}
  {2019})}\BibitemShut {NoStop}%
\bibitem [{\citenamefont {Horodecki}\ \emph {et~al.}(2009)\citenamefont
  {Horodecki}, \citenamefont {Horodecki}, \citenamefont {Horodecki},\ and\
  \citenamefont {Horodecki}}]{horodecki2009quantum}%
  \BibitemOpen
  \bibfield  {author} {\bibinfo {author} {\bibfnamefont {R.}~\bibnamefont
  {Horodecki}}, \bibinfo {author} {\bibfnamefont {P.}~\bibnamefont
  {Horodecki}}, \bibinfo {author} {\bibfnamefont {M.}~\bibnamefont
  {Horodecki}},\ and\ \bibinfo {author} {\bibfnamefont {K.}~\bibnamefont
  {Horodecki}},\ }\bibfield  {title} {\bibinfo {title} {Quantum entanglement},\
  }\href@noop {} {\bibfield  {journal} {\bibinfo  {journal} {Reviews of modern
  physics}\ }\textbf {\bibinfo {volume} {81}},\ \bibinfo {pages} {865}
  (\bibinfo {year} {2009})}\BibitemShut {NoStop}%
\bibitem [{\citenamefont {{\.Z}yczkowski}\ \emph {et~al.}(1998)\citenamefont
  {{\.Z}yczkowski}, \citenamefont {Horodecki}, \citenamefont {Sanpera},\ and\
  \citenamefont {Lewenstein}}]{zyczkowski1998volume}%
  \BibitemOpen
  \bibfield  {author} {\bibinfo {author} {\bibfnamefont {K.}~\bibnamefont
  {{\.Z}yczkowski}}, \bibinfo {author} {\bibfnamefont {P.}~\bibnamefont
  {Horodecki}}, \bibinfo {author} {\bibfnamefont {A.}~\bibnamefont {Sanpera}},\
  and\ \bibinfo {author} {\bibfnamefont {M.}~\bibnamefont {Lewenstein}},\
  }\bibfield  {title} {\bibinfo {title} {Volume of the set of separable
  states},\ }\href@noop {} {\bibfield  {journal} {\bibinfo  {journal} {Physical
  Review A}\ }\textbf {\bibinfo {volume} {58}},\ \bibinfo {pages} {883}
  (\bibinfo {year} {1998})}\BibitemShut {NoStop}%
\bibitem [{\citenamefont {Eisert}\ and\ \citenamefont
  {Plenio}(1999)}]{eisert1999comparison}%
  \BibitemOpen
  \bibfield  {author} {\bibinfo {author} {\bibfnamefont {J.}~\bibnamefont
  {Eisert}}\ and\ \bibinfo {author} {\bibfnamefont {M.~B.}\ \bibnamefont
  {Plenio}},\ }\bibfield  {title} {\bibinfo {title} {A comparison of
  entanglement measures},\ }\href@noop {} {\bibfield  {journal} {\bibinfo
  {journal} {Journal of Modern Optics}\ }\textbf {\bibinfo {volume} {46}},\
  \bibinfo {pages} {145} (\bibinfo {year} {1999})}\BibitemShut {NoStop}%
\bibitem [{\citenamefont {Daley}\ \emph {et~al.}(2012)\citenamefont {Daley},
  \citenamefont {Pichler}, \citenamefont {Schachenmayer},\ and\ \citenamefont
  {Zoller}}]{daley2012measuring}%
  \BibitemOpen
  \bibfield  {author} {\bibinfo {author} {\bibfnamefont {A.}~\bibnamefont
  {Daley}}, \bibinfo {author} {\bibfnamefont {H.}~\bibnamefont {Pichler}},
  \bibinfo {author} {\bibfnamefont {J.}~\bibnamefont {Schachenmayer}},\ and\
  \bibinfo {author} {\bibfnamefont {P.}~\bibnamefont {Zoller}},\ }\bibfield
  {title} {\bibinfo {title} {Measuring entanglement growth in quench dynamics
  of bosons in an optical lattice},\ }\href@noop {} {\bibfield  {journal}
  {\bibinfo  {journal} {Physical Review Letters}\ }\textbf {\bibinfo {volume}
  {109}},\ \bibinfo {pages} {020505} (\bibinfo {year} {2012})}\BibitemShut
  {NoStop}%
\bibitem [{\citenamefont {Abanin}\ and\ \citenamefont
  {Demler}(2012)}]{abanin2012measuring}%
  \BibitemOpen
  \bibfield  {author} {\bibinfo {author} {\bibfnamefont {D.~A.}\ \bibnamefont
  {Abanin}}\ and\ \bibinfo {author} {\bibfnamefont {E.}~\bibnamefont
  {Demler}},\ }\bibfield  {title} {\bibinfo {title} {Measuring entanglement
  entropy of a generic many-body system with a quantum switch},\ }\href@noop {}
  {\bibfield  {journal} {\bibinfo  {journal} {Physical Review Letters}\
  }\textbf {\bibinfo {volume} {109}},\ \bibinfo {pages} {020504} (\bibinfo
  {year} {2012})}\BibitemShut {NoStop}%
\bibitem [{\citenamefont {Brydges}\ \emph {et~al.}(2019)\citenamefont
  {Brydges}, \citenamefont {Elben}, \citenamefont {Jurcevic}, \citenamefont
  {Vermersch}, \citenamefont {Maier}, \citenamefont {Lanyon}, \citenamefont
  {Zoller}, \citenamefont {Blatt},\ and\ \citenamefont
  {Roos}}]{brydges2019probing}%
  \BibitemOpen
  \bibfield  {author} {\bibinfo {author} {\bibfnamefont {T.}~\bibnamefont
  {Brydges}}, \bibinfo {author} {\bibfnamefont {A.}~\bibnamefont {Elben}},
  \bibinfo {author} {\bibfnamefont {P.}~\bibnamefont {Jurcevic}}, \bibinfo
  {author} {\bibfnamefont {B.}~\bibnamefont {Vermersch}}, \bibinfo {author}
  {\bibfnamefont {C.}~\bibnamefont {Maier}}, \bibinfo {author} {\bibfnamefont
  {B.~P.}\ \bibnamefont {Lanyon}}, \bibinfo {author} {\bibfnamefont
  {P.}~\bibnamefont {Zoller}}, \bibinfo {author} {\bibfnamefont
  {R.}~\bibnamefont {Blatt}},\ and\ \bibinfo {author} {\bibfnamefont {C.~F.}\
  \bibnamefont {Roos}},\ }\bibfield  {title} {\bibinfo {title} {Probing
  r{\'e}nyi entanglement entropy via randomized measurements},\ }\href@noop {}
  {\bibfield  {journal} {\bibinfo  {journal} {Science}\ }\textbf {\bibinfo
  {volume} {364}},\ \bibinfo {pages} {260} (\bibinfo {year}
  {2019})}\BibitemShut {NoStop}%
\bibitem [{\citenamefont {Zhou}\ \emph {et~al.}(2020)\citenamefont {Zhou},
  \citenamefont {Zeng},\ and\ \citenamefont {Liu}}]{zhou2020single}%
  \BibitemOpen
  \bibfield  {author} {\bibinfo {author} {\bibfnamefont {Y.}~\bibnamefont
  {Zhou}}, \bibinfo {author} {\bibfnamefont {P.}~\bibnamefont {Zeng}},\ and\
  \bibinfo {author} {\bibfnamefont {Z.}~\bibnamefont {Liu}},\ }\bibfield
  {title} {\bibinfo {title} {Single-copies estimation of entanglement
  negativity},\ }\href@noop {} {\bibfield  {journal} {\bibinfo  {journal}
  {Physical Review Letters}\ }\textbf {\bibinfo {volume} {125}},\ \bibinfo
  {pages} {200502} (\bibinfo {year} {2020})}\BibitemShut {NoStop}%
\bibitem [{\citenamefont {Gray}\ \emph {et~al.}(2018)\citenamefont {Gray},
  \citenamefont {Banchi}, \citenamefont {Bayat},\ and\ \citenamefont
  {Bose}}]{gray2018machine}%
  \BibitemOpen
  \bibfield  {author} {\bibinfo {author} {\bibfnamefont {J.}~\bibnamefont
  {Gray}}, \bibinfo {author} {\bibfnamefont {L.}~\bibnamefont {Banchi}},
  \bibinfo {author} {\bibfnamefont {A.}~\bibnamefont {Bayat}},\ and\ \bibinfo
  {author} {\bibfnamefont {S.}~\bibnamefont {Bose}},\ }\bibfield  {title}
  {\bibinfo {title} {Machine-learning-assisted many-body entanglement
  measurement},\ }\href@noop {} {\bibfield  {journal} {\bibinfo  {journal}
  {Physical Review Letters}\ }\textbf {\bibinfo {volume} {121}},\ \bibinfo
  {pages} {150503} (\bibinfo {year} {2018})}\BibitemShut {NoStop}%
\bibitem [{\citenamefont {Berkovits}(2018)}]{berkovits2018extracting}%
  \BibitemOpen
  \bibfield  {author} {\bibinfo {author} {\bibfnamefont {R.}~\bibnamefont
  {Berkovits}},\ }\bibfield  {title} {\bibinfo {title} {Extracting
  many-particle entanglement entropy from observables using supervised machine
  learning},\ }\href@noop {} {\bibfield  {journal} {\bibinfo  {journal}
  {Physical Review B}\ }\textbf {\bibinfo {volume} {98}},\ \bibinfo {pages}
  {241411} (\bibinfo {year} {2018})}\BibitemShut {NoStop}%
\bibitem [{\citenamefont {Shahandeh}\ \emph {et~al.}(2017)\citenamefont
  {Shahandeh}, \citenamefont {Hall},\ and\ \citenamefont
  {Ralph}}]{shahandeh2017measurement}%
  \BibitemOpen
  \bibfield  {author} {\bibinfo {author} {\bibfnamefont {F.}~\bibnamefont
  {Shahandeh}}, \bibinfo {author} {\bibfnamefont {M.~J.}\ \bibnamefont
  {Hall}},\ and\ \bibinfo {author} {\bibfnamefont {T.~C.}\ \bibnamefont
  {Ralph}},\ }\bibfield  {title} {\bibinfo {title}
  {Measurement-device-independent approach to entanglement measures},\
  }\href@noop {} {\bibfield  {journal} {\bibinfo  {journal} {Physical Review
  Letters}\ }\textbf {\bibinfo {volume} {118}},\ \bibinfo {pages} {150505}
  (\bibinfo {year} {2017})}\BibitemShut {NoStop}%
\bibitem [{\citenamefont {Rosset}\ \emph {et~al.}(2018)\citenamefont {Rosset},
  \citenamefont {Martin}, \citenamefont {Verbanis}, \citenamefont {Lim},\ and\
  \citenamefont {Thew}}]{rosset2018practical}%
  \BibitemOpen
  \bibfield  {author} {\bibinfo {author} {\bibfnamefont {D.}~\bibnamefont
  {Rosset}}, \bibinfo {author} {\bibfnamefont {A.}~\bibnamefont {Martin}},
  \bibinfo {author} {\bibfnamefont {E.}~\bibnamefont {Verbanis}}, \bibinfo
  {author} {\bibfnamefont {C.~C.~W.}\ \bibnamefont {Lim}},\ and\ \bibinfo
  {author} {\bibfnamefont {R.}~\bibnamefont {Thew}},\ }\bibfield  {title}
  {\bibinfo {title} {Practical measurement-device-independent entanglement
  quantification},\ }\href@noop {} {\bibfield  {journal} {\bibinfo  {journal}
  {Physical Review A}\ }\textbf {\bibinfo {volume} {98}},\ \bibinfo {pages}
  {052332} (\bibinfo {year} {2018})}\BibitemShut {NoStop}%
\bibitem [{\citenamefont {Guo}\ \emph {et~al.}(2020)\citenamefont {Guo},
  \citenamefont {Yu}, \citenamefont {Hu}, \citenamefont {Liu}, \citenamefont
  {Wu}, \citenamefont {Huang}, \citenamefont {Li},\ and\ \citenamefont
  {Guo}}]{guo2020measurement}%
  \BibitemOpen
  \bibfield  {author} {\bibinfo {author} {\bibfnamefont {Y.}~\bibnamefont
  {Guo}}, \bibinfo {author} {\bibfnamefont {B.-C.}\ \bibnamefont {Yu}},
  \bibinfo {author} {\bibfnamefont {X.-M.}\ \bibnamefont {Hu}}, \bibinfo
  {author} {\bibfnamefont {B.-H.}\ \bibnamefont {Liu}}, \bibinfo {author}
  {\bibfnamefont {Y.-C.}\ \bibnamefont {Wu}}, \bibinfo {author} {\bibfnamefont
  {Y.-F.}\ \bibnamefont {Huang}}, \bibinfo {author} {\bibfnamefont {C.-F.}\
  \bibnamefont {Li}},\ and\ \bibinfo {author} {\bibfnamefont {G.-C.}\
  \bibnamefont {Guo}},\ }\bibfield  {title} {\bibinfo {title}
  {Measurement-device-independent quantification of irreducible
  high-dimensional entanglement},\ }\href@noop {} {\bibfield  {journal}
  {\bibinfo  {journal} {npj Quantum Information}\ }\textbf {\bibinfo {volume}
  {6}},\ \bibinfo {pages} {1} (\bibinfo {year} {2020})}\BibitemShut {NoStop}%
\bibitem [{\citenamefont {Moroder}\ \emph {et~al.}(2013)\citenamefont
  {Moroder}, \citenamefont {Bancal}, \citenamefont {Liang}, \citenamefont
  {Hofmann},\ and\ \citenamefont {G{\"u}hne}}]{moroder2013device}%
  \BibitemOpen
  \bibfield  {author} {\bibinfo {author} {\bibfnamefont {T.}~\bibnamefont
  {Moroder}}, \bibinfo {author} {\bibfnamefont {J.-D.}\ \bibnamefont {Bancal}},
  \bibinfo {author} {\bibfnamefont {Y.-C.}\ \bibnamefont {Liang}}, \bibinfo
  {author} {\bibfnamefont {M.}~\bibnamefont {Hofmann}},\ and\ \bibinfo {author}
  {\bibfnamefont {O.}~\bibnamefont {G{\"u}hne}},\ }\bibfield  {title} {\bibinfo
  {title} {Device-independent entanglement quantification and related
  applications},\ }\href@noop {} {\bibfield  {journal} {\bibinfo  {journal}
  {Physical Review Letters}\ }\textbf {\bibinfo {volume} {111}},\ \bibinfo
  {pages} {030501} (\bibinfo {year} {2013})}\BibitemShut {NoStop}%
\bibitem [{\citenamefont {Bardyn}\ \emph {et~al.}(2009)\citenamefont {Bardyn},
  \citenamefont {Liew}, \citenamefont {Massar}, \citenamefont {McKague},\ and\
  \citenamefont {Scarani}}]{bardyn2009device}%
  \BibitemOpen
  \bibfield  {author} {\bibinfo {author} {\bibfnamefont {C.-E.}\ \bibnamefont
  {Bardyn}}, \bibinfo {author} {\bibfnamefont {T.~C.}\ \bibnamefont {Liew}},
  \bibinfo {author} {\bibfnamefont {S.}~\bibnamefont {Massar}}, \bibinfo
  {author} {\bibfnamefont {M.}~\bibnamefont {McKague}},\ and\ \bibinfo {author}
  {\bibfnamefont {V.}~\bibnamefont {Scarani}},\ }\bibfield  {title} {\bibinfo
  {title} {Device-independent state estimation based on bell's inequalities},\
  }\href@noop {} {\bibfield  {journal} {\bibinfo  {journal} {Physical Review
  A}\ }\textbf {\bibinfo {volume} {80}},\ \bibinfo {pages} {062327} (\bibinfo
  {year} {2009})}\BibitemShut {NoStop}%
\bibitem [{\citenamefont {Kaniewski}(2016)}]{kaniewski2016analytic}%
  \BibitemOpen
  \bibfield  {author} {\bibinfo {author} {\bibfnamefont {J.}~\bibnamefont
  {Kaniewski}},\ }\bibfield  {title} {\bibinfo {title} {Analytic and nearly
  optimal self-testing bounds for the clauser-horne-shimony-holt and mermin
  inequalities},\ }\href@noop {} {\bibfield  {journal} {\bibinfo  {journal}
  {Physical Review Letters}\ }\textbf {\bibinfo {volume} {117}},\ \bibinfo
  {pages} {070402} (\bibinfo {year} {2016})}\BibitemShut {NoStop}%
\bibitem [{\citenamefont {Wei}\ and\ \citenamefont
  {Lin}(2021)}]{wei2021analytic}%
  \BibitemOpen
  \bibfield  {author} {\bibinfo {author} {\bibfnamefont {Z.}~\bibnamefont
  {Wei}}\ and\ \bibinfo {author} {\bibfnamefont {L.}~\bibnamefont {Lin}},\
  }\bibfield  {title} {\bibinfo {title} {Analytic semi-device-independent
  entanglement quantification for bipartite quantum states},\ }\href@noop {}
  {\bibfield  {journal} {\bibinfo  {journal} {Physical Review A}\ }\textbf
  {\bibinfo {volume} {103}},\ \bibinfo {pages} {032215} (\bibinfo {year}
  {2021})}\BibitemShut {NoStop}%
\bibitem [{\citenamefont {Cornelio}\ \emph {et~al.}(2011)\citenamefont
  {Cornelio}, \citenamefont {de~Oliveira},\ and\ \citenamefont
  {Fanchini}}]{cornelio2011entanglement}%
  \BibitemOpen
  \bibfield  {author} {\bibinfo {author} {\bibfnamefont {M.~F.}\ \bibnamefont
  {Cornelio}}, \bibinfo {author} {\bibfnamefont {M.~C.}\ \bibnamefont
  {de~Oliveira}},\ and\ \bibinfo {author} {\bibfnamefont {F.~F.}\ \bibnamefont
  {Fanchini}},\ }\bibfield  {title} {\bibinfo {title} {Entanglement
  irreversibility from quantum discord and quantum deficit},\ }\href@noop {}
  {\bibfield  {journal} {\bibinfo  {journal} {Physical Review Letters}\
  }\textbf {\bibinfo {volume} {107}},\ \bibinfo {pages} {020502} (\bibinfo
  {year} {2011})}\BibitemShut {NoStop}%
\bibitem [{\citenamefont {Lin}\ and\ \citenamefont
  {Wei}(2020)}]{lin2020quantifying}%
  \BibitemOpen
  \bibfield  {author} {\bibinfo {author} {\bibfnamefont {L.}~\bibnamefont
  {Lin}}\ and\ \bibinfo {author} {\bibfnamefont {Z.}~\bibnamefont {Wei}},\
  }\bibfield  {title} {\bibinfo {title} {Quantifying multipartite quantum
  entanglement by nondegenerate bell inequalities},\ }\href@noop {} {\bibfield
  {journal} {\bibinfo  {journal} {arXiv preprint arXiv:2008.12064}\ } (\bibinfo
  {year} {2020})}\BibitemShut {NoStop}%
\bibitem [{\citenamefont {Schumacher}\ and\ \citenamefont
  {Nielsen}(1996)}]{schumacher1996quantum}%
  \BibitemOpen
  \bibfield  {author} {\bibinfo {author} {\bibfnamefont {B.}~\bibnamefont
  {Schumacher}}\ and\ \bibinfo {author} {\bibfnamefont {M.~A.}\ \bibnamefont
  {Nielsen}},\ }\bibfield  {title} {\bibinfo {title} {Quantum data processing
  and error correction},\ }\href@noop {} {\bibfield  {journal} {\bibinfo
  {journal} {Physical Review A}\ }\textbf {\bibinfo {volume} {54}},\ \bibinfo
  {pages} {2629} (\bibinfo {year} {1996})}\BibitemShut {NoStop}%
\bibitem [{\citenamefont {Lloyd}(1997)}]{lloyd1997capacity}%
  \BibitemOpen
  \bibfield  {author} {\bibinfo {author} {\bibfnamefont {S.}~\bibnamefont
  {Lloyd}},\ }\bibfield  {title} {\bibinfo {title} {Capacity of the noisy
  quantum channel},\ }\href@noop {} {\bibfield  {journal} {\bibinfo  {journal}
  {Physical Review A}\ }\textbf {\bibinfo {volume} {55}},\ \bibinfo {pages}
  {1613} (\bibinfo {year} {1997})}\BibitemShut {NoStop}%
\bibitem [{\citenamefont {Shimony}(1995)}]{shimony1995degree}%
  \BibitemOpen
  \bibfield  {author} {\bibinfo {author} {\bibfnamefont {A.}~\bibnamefont
  {Shimony}},\ }\bibfield  {title} {\bibinfo {title} {Degree of entanglement
  a},\ }\href@noop {} {\bibfield  {journal} {\bibinfo  {journal} {Annals of the
  New York Academy of Sciences}\ }\textbf {\bibinfo {volume} {755}},\ \bibinfo
  {pages} {675} (\bibinfo {year} {1995})}\BibitemShut {NoStop}%
\bibitem [{\citenamefont {Barnum}\ and\ \citenamefont
  {Linden}(2001)}]{barnum2001monotones}%
  \BibitemOpen
  \bibfield  {author} {\bibinfo {author} {\bibfnamefont {H.}~\bibnamefont
  {Barnum}}\ and\ \bibinfo {author} {\bibfnamefont {N.}~\bibnamefont
  {Linden}},\ }\bibfield  {title} {\bibinfo {title} {Monotones and invariants
  for multi-particle quantum states},\ }\href@noop {} {\bibfield  {journal}
  {\bibinfo  {journal} {Journal of Physics A: Mathematical and General}\
  }\textbf {\bibinfo {volume} {34}},\ \bibinfo {pages} {6787} (\bibinfo {year}
  {2001})}\BibitemShut {NoStop}%
\bibitem [{\citenamefont {Collins}\ \emph
  {et~al.}(2002{\natexlab{b}})\citenamefont {Collins}, \citenamefont {Gisin},
  \citenamefont {Linden}, \citenamefont {Massar},\ and\ \citenamefont
  {Popescu}}]{collins2002bell}%
  \BibitemOpen
  \bibfield  {author} {\bibinfo {author} {\bibfnamefont {D.}~\bibnamefont
  {Collins}}, \bibinfo {author} {\bibfnamefont {N.}~\bibnamefont {Gisin}},
  \bibinfo {author} {\bibfnamefont {N.}~\bibnamefont {Linden}}, \bibinfo
  {author} {\bibfnamefont {S.}~\bibnamefont {Massar}},\ and\ \bibinfo {author}
  {\bibfnamefont {S.}~\bibnamefont {Popescu}},\ }\bibfield  {title} {\bibinfo
  {title} {Bell inequalities for arbitrarily high-dimensional systems},\
  }\href@noop {} {\bibfield  {journal} {\bibinfo  {journal} {Physical Review
  Letters}\ }\textbf {\bibinfo {volume} {88}},\ \bibinfo {pages} {040404}
  (\bibinfo {year} {2002}{\natexlab{b}})}\BibitemShut {NoStop}%
\bibitem [{\citenamefont {Acin}\ \emph {et~al.}(2002)\citenamefont {Acin},
  \citenamefont {Durt}, \citenamefont {Gisin},\ and\ \citenamefont
  {Latorre}}]{acin2002quantum}%
  \BibitemOpen
  \bibfield  {author} {\bibinfo {author} {\bibfnamefont {A.}~\bibnamefont
  {Acin}}, \bibinfo {author} {\bibfnamefont {T.}~\bibnamefont {Durt}}, \bibinfo
  {author} {\bibfnamefont {N.}~\bibnamefont {Gisin}},\ and\ \bibinfo {author}
  {\bibfnamefont {J.~I.}\ \bibnamefont {Latorre}},\ }\bibfield  {title}
  {\bibinfo {title} {Quantum nonlocality in two three-level systems},\
  }\href@noop {} {\bibfield  {journal} {\bibinfo  {journal} {Physical Review
  A}\ }\textbf {\bibinfo {volume} {65}},\ \bibinfo {pages} {052325} (\bibinfo
  {year} {2002})}\BibitemShut {NoStop}%
\bibitem [{\citenamefont {Chen}\ \emph {et~al.}(2006)\citenamefont {Chen},
  \citenamefont {Wu}, \citenamefont {Kwek}, \citenamefont {Oh},\ and\
  \citenamefont {Ge}}]{chen2006violating}%
  \BibitemOpen
  \bibfield  {author} {\bibinfo {author} {\bibfnamefont {J.-L.}\ \bibnamefont
  {Chen}}, \bibinfo {author} {\bibfnamefont {C.}~\bibnamefont {Wu}}, \bibinfo
  {author} {\bibfnamefont {L.~C.}\ \bibnamefont {Kwek}}, \bibinfo {author}
  {\bibfnamefont {C.~H.}\ \bibnamefont {Oh}},\ and\ \bibinfo {author}
  {\bibfnamefont {M.-L.}\ \bibnamefont {Ge}},\ }\bibfield  {title} {\bibinfo
  {title} {Violating bell inequalities maximally for two d-dimensional
  systems},\ }\href@noop {} {\bibfield  {journal} {\bibinfo  {journal}
  {Physical Review A}\ }\textbf {\bibinfo {volume} {74}},\ \bibinfo {pages}
  {032106} (\bibinfo {year} {2006})}\BibitemShut {NoStop}%
\bibitem [{\citenamefont {Zohren}\ and\ \citenamefont
  {Gill}(2008)}]{zohren2008maximal}%
  \BibitemOpen
  \bibfield  {author} {\bibinfo {author} {\bibfnamefont {S.}~\bibnamefont
  {Zohren}}\ and\ \bibinfo {author} {\bibfnamefont {R.~D.}\ \bibnamefont
  {Gill}},\ }\bibfield  {title} {\bibinfo {title} {Maximal violation of the
  collins-gisin-linden-massar-popescu inequality for infinite dimensional
  states},\ }\href@noop {} {\bibfield  {journal} {\bibinfo  {journal} {Physical
  Review Letters}\ }\textbf {\bibinfo {volume} {100}},\ \bibinfo {pages}
  {120406} (\bibinfo {year} {2008})}\BibitemShut {NoStop}%
\bibitem [{\citenamefont {Barrett}\ \emph {et~al.}(2006)\citenamefont
  {Barrett}, \citenamefont {Kent},\ and\ \citenamefont
  {Pironio}}]{barrett2006maximally}%
  \BibitemOpen
  \bibfield  {author} {\bibinfo {author} {\bibfnamefont {J.}~\bibnamefont
  {Barrett}}, \bibinfo {author} {\bibfnamefont {A.}~\bibnamefont {Kent}},\ and\
  \bibinfo {author} {\bibfnamefont {S.}~\bibnamefont {Pironio}},\ }\bibfield
  {title} {\bibinfo {title} {Maximally nonlocal and monogamous quantum
  correlations},\ }\href@noop {} {\bibfield  {journal} {\bibinfo  {journal}
  {Physical Review Letters}\ }\textbf {\bibinfo {volume} {97}},\ \bibinfo
  {pages} {170409} (\bibinfo {year} {2006})}\BibitemShut {NoStop}%
\bibitem [{\citenamefont {LeCun}\ \emph {et~al.}(1998)\citenamefont {LeCun},
  \citenamefont {Bottou}, \citenamefont {Bengio},\ and\ \citenamefont
  {Haffner}}]{lecun1998gradient}%
  \BibitemOpen
  \bibfield  {author} {\bibinfo {author} {\bibfnamefont {Y.}~\bibnamefont
  {LeCun}}, \bibinfo {author} {\bibfnamefont {L.}~\bibnamefont {Bottou}},
  \bibinfo {author} {\bibfnamefont {Y.}~\bibnamefont {Bengio}},\ and\ \bibinfo
  {author} {\bibfnamefont {P.}~\bibnamefont {Haffner}},\ }\bibfield  {title}
  {\bibinfo {title} {Gradient-based learning applied to document recognition},\
  }\href@noop {} {\bibfield  {journal} {\bibinfo  {journal} {Proceedings of the
  IEEE}\ }\textbf {\bibinfo {volume} {86}},\ \bibinfo {pages} {2278} (\bibinfo
  {year} {1998})}\BibitemShut {NoStop}%
\bibitem [{\citenamefont {Albawi}\ \emph {et~al.}(2017)\citenamefont {Albawi},
  \citenamefont {Mohammed},\ and\ \citenamefont
  {Al-Zawi}}]{albawi2017understanding}%
  \BibitemOpen
  \bibfield  {author} {\bibinfo {author} {\bibfnamefont {S.}~\bibnamefont
  {Albawi}}, \bibinfo {author} {\bibfnamefont {T.~A.}\ \bibnamefont
  {Mohammed}},\ and\ \bibinfo {author} {\bibfnamefont {S.}~\bibnamefont
  {Al-Zawi}},\ }\bibfield  {title} {\bibinfo {title} {Understanding of a
  convolutional neural network},\ }in\ \href@noop {} {\emph {\bibinfo
  {booktitle} {2017 International Conference on Engineering and Technology
  (ICET)}}}\ (\bibinfo {organization} {Ieee},\ \bibinfo {year} {2017})\ pp.\
  \bibinfo {pages} {1--6}\BibitemShut {NoStop}%
\bibitem [{\citenamefont {D{\"u}r}\ \emph {et~al.}(2000)\citenamefont
  {D{\"u}r}, \citenamefont {Vidal},\ and\ \citenamefont
  {Cirac}}]{dur2000three}%
  \BibitemOpen
  \bibfield  {author} {\bibinfo {author} {\bibfnamefont {W.}~\bibnamefont
  {D{\"u}r}}, \bibinfo {author} {\bibfnamefont {G.}~\bibnamefont {Vidal}},\
  and\ \bibinfo {author} {\bibfnamefont {J.~I.}\ \bibnamefont {Cirac}},\
  }\bibfield  {title} {\bibinfo {title} {Three qubits can be entangled in two
  inequivalent ways},\ }\href@noop {} {\bibfield  {journal} {\bibinfo
  {journal} {Physical Review A}\ }\textbf {\bibinfo {volume} {62}},\ \bibinfo
  {pages} {062314} (\bibinfo {year} {2000})}\BibitemShut {NoStop}%
\bibitem [{\citenamefont {Greenberger}\ \emph {et~al.}(1989)\citenamefont
  {Greenberger}, \citenamefont {Horne},\ and\ \citenamefont
  {Zeilinger}}]{greenberger1989going}%
  \BibitemOpen
  \bibfield  {author} {\bibinfo {author} {\bibfnamefont {D.~M.}\ \bibnamefont
  {Greenberger}}, \bibinfo {author} {\bibfnamefont {M.~A.}\ \bibnamefont
  {Horne}},\ and\ \bibinfo {author} {\bibfnamefont {A.}~\bibnamefont
  {Zeilinger}},\ }\bibfield  {title} {\bibinfo {title} {Going beyond bell's
  theorem},\ }in\ \href@noop {} {\emph {\bibinfo {booktitle} {Bell's theorem,
  quantum theory and conceptions of the universe}}}\ (\bibinfo  {publisher}
  {Springer},\ \bibinfo {year} {1989})\ pp.\ \bibinfo {pages}
  {69--72}\BibitemShut {NoStop}%
\bibitem [{\citenamefont {Hensen}\ \emph {et~al.}(2015)\citenamefont {Hensen},
  \citenamefont {Bernien}, \citenamefont {Dr{\'e}au}, \citenamefont {Reiserer},
  \citenamefont {Kalb}, \citenamefont {Blok}, \citenamefont {Ruitenberg},
  \citenamefont {Vermeulen}, \citenamefont {Schouten}, \citenamefont
  {Abell{\'a}n} \emph {et~al.}}]{hensen2015loophole}%
  \BibitemOpen
  \bibfield  {author} {\bibinfo {author} {\bibfnamefont {B.}~\bibnamefont
  {Hensen}}, \bibinfo {author} {\bibfnamefont {H.}~\bibnamefont {Bernien}},
  \bibinfo {author} {\bibfnamefont {A.~E.}\ \bibnamefont {Dr{\'e}au}}, \bibinfo
  {author} {\bibfnamefont {A.}~\bibnamefont {Reiserer}}, \bibinfo {author}
  {\bibfnamefont {N.}~\bibnamefont {Kalb}}, \bibinfo {author} {\bibfnamefont
  {M.~S.}\ \bibnamefont {Blok}}, \bibinfo {author} {\bibfnamefont
  {J.}~\bibnamefont {Ruitenberg}}, \bibinfo {author} {\bibfnamefont {R.~F.}\
  \bibnamefont {Vermeulen}}, \bibinfo {author} {\bibfnamefont {R.~N.}\
  \bibnamefont {Schouten}}, \bibinfo {author} {\bibfnamefont {C.}~\bibnamefont
  {Abell{\'a}n}}, \emph {et~al.},\ }\bibfield  {title} {\bibinfo {title}
  {Loophole-free bell inequality violation using electron spins separated by
  1.3 kilometres},\ }\href@noop {} {\bibfield  {journal} {\bibinfo  {journal}
  {Nature}\ }\textbf {\bibinfo {volume} {526}},\ \bibinfo {pages} {682}
  (\bibinfo {year} {2015})}\BibitemShut {NoStop}%
\bibitem [{\citenamefont {Giustina}\ \emph {et~al.}(2015)\citenamefont
  {Giustina}, \citenamefont {Versteegh}, \citenamefont {Wengerowsky},
  \citenamefont {Handsteiner}, \citenamefont {Hochrainer}, \citenamefont
  {Phelan}, \citenamefont {Steinlechner}, \citenamefont {Kofler}, \citenamefont
  {Larsson}, \citenamefont {Abell{\'a}n} \emph
  {et~al.}}]{giustina2015significant}%
  \BibitemOpen
  \bibfield  {author} {\bibinfo {author} {\bibfnamefont {M.}~\bibnamefont
  {Giustina}}, \bibinfo {author} {\bibfnamefont {M.~A.}\ \bibnamefont
  {Versteegh}}, \bibinfo {author} {\bibfnamefont {S.}~\bibnamefont
  {Wengerowsky}}, \bibinfo {author} {\bibfnamefont {J.}~\bibnamefont
  {Handsteiner}}, \bibinfo {author} {\bibfnamefont {A.}~\bibnamefont
  {Hochrainer}}, \bibinfo {author} {\bibfnamefont {K.}~\bibnamefont {Phelan}},
  \bibinfo {author} {\bibfnamefont {F.}~\bibnamefont {Steinlechner}}, \bibinfo
  {author} {\bibfnamefont {J.}~\bibnamefont {Kofler}}, \bibinfo {author}
  {\bibfnamefont {J.-{\AA}.}\ \bibnamefont {Larsson}}, \bibinfo {author}
  {\bibfnamefont {C.}~\bibnamefont {Abell{\'a}n}}, \emph {et~al.},\ }\bibfield
  {title} {\bibinfo {title} {Significant-loophole-free test of bell's theorem
  with entangled photons},\ }\href@noop {} {\bibfield  {journal} {\bibinfo
  {journal} {Physical Review Letters}\ }\textbf {\bibinfo {volume} {115}},\
  \bibinfo {pages} {250401} (\bibinfo {year} {2015})}\BibitemShut {NoStop}%
\bibitem [{\citenamefont {Shalm}\ \emph {et~al.}(2015)\citenamefont {Shalm},
  \citenamefont {Meyer-Scott}, \citenamefont {Christensen}, \citenamefont
  {Bierhorst}, \citenamefont {Wayne}, \citenamefont {Stevens}, \citenamefont
  {Gerrits}, \citenamefont {Glancy}, \citenamefont {Hamel}, \citenamefont
  {Allman} \emph {et~al.}}]{shalm2015strong}%
  \BibitemOpen
  \bibfield  {author} {\bibinfo {author} {\bibfnamefont {L.~K.}\ \bibnamefont
  {Shalm}}, \bibinfo {author} {\bibfnamefont {E.}~\bibnamefont {Meyer-Scott}},
  \bibinfo {author} {\bibfnamefont {B.~G.}\ \bibnamefont {Christensen}},
  \bibinfo {author} {\bibfnamefont {P.}~\bibnamefont {Bierhorst}}, \bibinfo
  {author} {\bibfnamefont {M.~A.}\ \bibnamefont {Wayne}}, \bibinfo {author}
  {\bibfnamefont {M.~J.}\ \bibnamefont {Stevens}}, \bibinfo {author}
  {\bibfnamefont {T.}~\bibnamefont {Gerrits}}, \bibinfo {author} {\bibfnamefont
  {S.}~\bibnamefont {Glancy}}, \bibinfo {author} {\bibfnamefont {D.~R.}\
  \bibnamefont {Hamel}}, \bibinfo {author} {\bibfnamefont {M.~S.}\ \bibnamefont
  {Allman}}, \emph {et~al.},\ }\bibfield  {title} {\bibinfo {title} {Strong
  loophole-free test of local realism},\ }\href@noop {} {\bibfield  {journal}
  {\bibinfo  {journal} {Physical Review Letters}\ }\textbf {\bibinfo {volume}
  {115}},\ \bibinfo {pages} {250402} (\bibinfo {year} {2015})}\BibitemShut
  {NoStop}%
\end{thebibliography}%

\end{document}